\documentclass{article}
\usepackage[english]{babel}
\usepackage[letterpaper,top=2cm,bottom=2cm,left=3cm,right=3cm,marginparwidth=1.75cm]{geometry}

\usepackage{amsmath}
\usepackage{amssymb}
\usepackage{graphicx}
\usepackage{booktabs}
\usepackage{subcaption}
\usepackage{xcolor}
\usepackage{authblk}

\usepackage{natbib}

\usepackage{lipsum}

\usepackage{blindtext}

\newcommand{\Tb}{M} 
\newcommand{\ib}{m} 
\newcommand{\Tc}{N} 
\newcommand{\ic}{n} 
\newcommand{\Tcp}{K} 
\newcommand{\bt}{\boldsymbol{\theta}} 
\newcommand{\icluster}{r} 
\newcommand{\Tcluster}{d} 

\newcommand{\pck}[1]{\textsf{#1}}

\title{Bayesian Clustering of Constant-Wise Change-Point Data}
\author{Ana Carolina da Cruz$^{1*}$ \and \ Camila P.\ E.\ de Souza$^1$\\  $^1$Department of Statistical and Actuarial Sciences, University of Western Ontario
$^*$Corresponding author: adacruz@uwo.ca}

\date{}

\begin{document}
\maketitle

\begin{abstract}

Change-point models deal with ordered data sequences. Their primary goal is to infer the locations where an aspect of the data sequence changes. In this paper, we propose and implement a nonparametric Bayesian model for clustering observations based on their constant-wise change-point profiles via Gibbs sampler. Our model incorporates a Dirichlet Process on the constant-wise change-point structures to cluster observations while simultaneously performing multiple change-point estimation. Additionally, our approach controls the number of clusters in the model, not requiring the specification of the number of clusters \textit{a priori}. Satisfactory clustering and estimation results were obtained when evaluating our method under various simulated scenarios and on a real dataset from single-cell genomic sequencing. Our proposed methodology is implemented as an R package called BayesCPclust and is available from the Comprehensive R Archive Network at \texttt{https://CRAN.R-project.org/package=BayesCPclust}.


\end{abstract}

\section{Introduction}

Change-point models deal with the analysis of an ordered sequence of random quantities. Examples of such sequences include daily average temperatures over time and sequencing data in genomics. An important component of a change-point model is change-point detection, which involves inferring the positions where an aspect of the data sequence changes, such as location or distribution. These change points and their corresponding locations are of great practical interest. One of the first applications of these models dates back to the 1950s when \cite{Page1954, Page1955} introduced a now well-know sequential method called cumulative sum (CUSUM) to detect changes in the mean of a quality control process. Since then, change-point detection has been actively addressed in various application settings, such as financial analysis \citep{Fryzlewicz2014} and biostatistics \citep{Olshen2004, Picard2011, Hocking2013}. Change-point detection is also widely studied in time-series analysis \citep{Jandhyala2013, Aue2013, Yan2019, Zhao2019, Militino2020}; however, in what follows, we focus on non-time-series techniques.

Change-point models are generally divided into two main groups: online methods, which perform sequential detection with new data continually arriving, commonly used in anomaly detection, and offline methods, in which retrospective analysis is performed in the entire observed sequence \citep{Truong2020}. In this article, we focus on the latter. Additionally, change-point models may be either parametric or nonparametric. Parametric models assume that the underlying distributions belong to some known family. In contrast, nonparametric approaches heavily rely on the estimation of density functions, but may be employed in a broader range of applications \citep{Brodsky1993, Chen2012, Haynes2017, Londschien2023}.

The literature on change-point models is vast, and several methods to perform change-point detection have been proposed in the past few decades, so we discuss here some approaches proposed for single and multiple change-point problems. For example, \cite{Olshen2004} and \cite{Fryzlewicz2014} proposed circular binary segmentation and wild binary segmentation, respectively, both based on the binary segmentation algorithm proposed by \cite{Vostrikova1981}. These methods perform change-point tests sequentially to locate change points in the data sequence. Other methods, mainly used for multiple change-point problems, treat change-point detection as a model-selection problem and estimate change points via minimizing a criterion. These methods often require dynamic programming such as the pruned exactly linear time (PELT) algorithm \citep{Killick2012} and the functional pruned (FP) algorithm \citep{Rigaill2015}. Some well-known approaches for multiple change-point detection include the simultaneous multiscale change-point estimator (SMUCE) \citep{Frick2014} and the heterogeneous simultaneous multiscale change-point estimator (H-SMUCE) \citep{Pein2017}, both based on a multiscale hypothesis testing, where the optimization process relies on the penalization of a test statistic. Additional approaches for change-point problems are described in \cite{Zou2014} and \cite{Niu2016}.

While all previously described approaches have been proposed to detect change points from a unique sequence of data points, there needs to be more literature on clustering change-point data from multiple sequences, especially considering only model-based techniques. To our knowledge, techniques involving change-point estimation and model-based clustering have only been studied by \cite{Dass2015}, \cite{Zhu2022}, and \cite{Sarkar2022}.
\cite{Zhu2022} proposed a finite Gaussian mixture model for clustering observations with a single change point, whereas \cite{Sarkar2022} proposed a finite Negative Binomial mixture model for clustering multiple change-point data. Both approaches consider the Expectation and Maximization (EM) algorithm for estimating the cluster assignments and the model parameters. 
The single change-point detection in \cite{Zhu2022} is performed using exhaustively searches for changes in the mean or variance where competing models are compared based on the Bayesian information criterion (BIC). 
The multiple change-point approach detects changes in the mean of a count process by employing a combination of segmentation and an exhaustive search approach. Similar to the single change-point approach, the best model is selected based on the BIC. Focusing on the analysis of the mortality rate over time for 49 states in the United States, \cite{Dass2015} took a Bayesian approach for clustering multiple change-point data by assuming a Functional Dirichlet Process on the linear piece-wise structure of their data to cluster states based on the change-point locations and slope magnitudes. Although these papers showed promising results in dealing with the problem of clustering change-point data while simultaneously performing change-point detection, none made their algorithm's implementation available. Moreover, there is currently no available software in R that simultaneously performs clustering and multiple change-point detection. Existing packages like \pck{ecp} \citep{james2013} and \pck{bcp} \citep{erdman2008} can detect multiple change points within a single sequence of observations but do not perform clustering over multiple sequences.

Bayesian techniques for change-point detection are known to provide state-of-the-art results in various settings, as stated in \cite{Dass2015, Truong2020}. We propose and implement as a R Package a nonparametric Bayesian model for clustering multiple constant-wise change-point data via Gibbs sampler. Similarly to \cite{Dass2015}, our model incorporates a Functional Dirichlet Process on the constant-wise change-point structures that automatically controls the number of clusters in the model as opposed to other clustering techniques \citep{Neal1992, Yerebakan2014}. To the best of our knowledge, this is the first paper to provide an implementation for the problem of clustering multiple change-point data while simultaneously performing change-point detection. We apply our proposed approach to cluster abnormal (tumour) single-cell genomic data based on their copy-number profiles, which resemble constant-wise structures. In addition, we evaluate the performance of our method under various simulated scenarios. Our proposed method is implemented as an R package called BayesCPclust and is available from the Comprehensive R Archive Network at \texttt{https://CRAN.R-project.org/package=BayesCPclust}.

The rest of the paper is organized as follows. Section 2 introduces our proposed methodology and provides the updating steps for the Gibbs sampler. Section 3 presents the performance results for our proposed method under various simulated scenarios. Finally, in Section 4, we show the application results of our method in a single-cell copy-number dataset.

\section{Methods}

Let $\mathbf{Y}_\ic = (Y_{\ic1}, \dotsc,  Y_{\ic\Tb})$ be a data sequence ordered based on some covariate such as time or position along a chromosome. For example, in the copy-number dataset in Section \ref{sec:real_data}, $Y_{\ic\ib}$ represents the $\log 2$ ratio GC-corrected copy number aligned to genomic bin $\ib$ and cell $\ic$, where $\ic = 1, \dotsc, \Tc$, and $\ib = 1, \dotsc, \Tb$.

If we assume that there are $\Tcp_\ic$ change points in $\mathbf{Y}_\ic$, that means that $\mathbf{Y}_\ic$ can be partitioned into $\Tcp_\ic + 1$ distinct segments, $[1, \tau^{(\ic)}_1), [\tau^{(\ic)}_1, \tau^{(\ic)}_2), \dotsc, [\tau^{(\ic)}_{\Tcp_\ic}, \Tb]$, with change-point positions $\tau^{(\ic)}_1, \dotsc, \tau^{(\ic)}_{\Tcp_\ic}$, such that $\tau^{(\ic)}_0 = 1$ and $\tau^{(\ic)}_{{\Tcp_\ic} + 1} = \Tb$. Also we assume that the change points are ordered, that is, $\tau^{(\ic)}_i < \tau^{(\ic)}_j$ if, and only if, $i < j$.

In our approach we assume a constant-wise structure for $Y_{\ic\ib}$ defined by the following the model:

\begin{equation}
\label{eq:model}
   Y_{\ic\ib} = \alpha_l^{(\ic)} + \epsilon_{\ic\ib},
\end{equation}
where $ \ib \in [\tau_{l-1}^{(\ic)}, \tau_l^{(\ic)})$ for $l = 1, \dotsc, \Tcp_\ic + 1$ and $\epsilon_{\ic\ib} \sim N(0, \sigma_\ic^2)$.

The model in Equation (\ref{eq:model}) assumes that the mean in each interval between change points is constant, defined by an intercept $\alpha_l$, $l = 1, ... ,  \Tcp_\ic + 1$. Furthermore, this model allows the variability around the mean to differ depending on the observation by specifying a variance component $\sigma_\ic^2$ for each $\ic$. 

Clustering change-point data via a Functional Dirichlet Process is formulated by assuming that the constant-wise structures for the observations are independent draws from some distribution, $G$, which in turn follows a Dirichlet Process prior. We define the constant-wise function as:

\begin{equation*}
    \bt_\ic(\ib) = \alpha_l^{(\ic)}, \qquad \text{if} \quad \tau_{l-1}^{(\ic)} \leq \ib \leq \tau_l^{(\ic)} - 1,
\end{equation*}
where $\alpha_l^{(\ic)}$ is the intercept in the segment $[\tau_{l-1}^{(\ic)}, \tau_l^{(\ic)})$ for each observation $\ic$. This constant-wise function, $\bt_\ic(\ib)$, contains all information about the number of change points, their locations and the intercepts for the corresponding segment.
Furthermore, a Dirichlet Process on $\bt_\ic$ leads to the hierarchical model:

\begin{align*}
    Y_{\ic\ib} \mid \bt_\ic, \sigma^2_\ic &\sim N(\alpha_l^{(\ic)}, \sigma^2_\ic),\\ \nonumber
    \bt_\ic \mid G &\sim G, \\
    G &\sim DP(\alpha_0, G_0), \nonumber
\end{align*}
where $G_0$ is the baseline distribution, such that $E(G) = G_0$, and $\alpha_0$ is the precision parameter that determines how distant the distribution $G \sim DP(\alpha_0, G_0)$ is from $G_0$. 

Integration over $G$ allows the predictive distribution of $\bt_\ic$ to be written as \citep{Blackwell1973}:

\begin{equation}
\label{eq:DP}
    \bt_\ic \mid \bt_{-\ic} \sim \frac{1}{N - 1 + \alpha_0}\sum_{j \neq \ic}\delta(\bt_j) + \frac{\alpha_0}{N - 1 + \alpha_0}G_0,
\end{equation}
where $\delta(\bt_j)$ is a point mass distribution at $\bt_j$, and $-\ic$ represents all the observations except for $\ic$.
Note that, by the first term in Equation (\ref{eq:DP}), there is a positive probability that draws from $G$ will take on the same value. It implies that for a long enough sequence of draws from $G$, the value of any draw will be repeated by another draw, indicating that $G$ is a discrete distribution. Therefore, a  Dirichlet Process on the change-point structures allows the proposed approach to control the number of clusters in the model, not requiring pre-specification. More details about the Dirichlet Process can be found in \cite{Neal2000} and \cite{Li2019}.

We define the distribution $G_0$ in the following hierarchical form to cluster observations according to their constant-wise change-point profiles.

\begin{itemize}
    \item[(i)] Distribution of the number of change points ($\Tcp$):  We assume that each segment between change points has at least $w > 0$ points to ensure non-zero length. Let $m_l$ be the interval length of the $l$-th segment after subtracting $w$. As a result, $\Tcp \leq k^*$, where $$k^* = \frac{\Tb - 1}{w} - 1$$ to unsure that $m_0 = \sum_{l = 1}^{\Tcp + 1}m_l = \Tb - 1 - (\Tcp + 1)w > 0$.

    Therefore, $\Tcp$ follows a truncated Poisson distribution given by:
    
    \begin{equation*}
        P(\Tcp = k) = \frac{e^{-\lambda}\lambda^k/k!}{\displaystyle\sum_{l=0}^{k^*}e^{-\lambda}\lambda^l/l!}, \quad \text{for } k = 0, \dotsc, k^*.
    \end{equation*}
    
    \item[(ii)] Distribution of the interval lengths between change points: Given $\Tcp = k$, the distribution of the interval lengths is defined as:
    
    \begin{equation*}
        (m_1, \dotsc, m_{k+1}) \mid \Tcp = k \sim \mathrm{Multinomial}\left(m_0, \frac{1}{k+1}, \dotsc, \frac{1}{k+1}\right).
    \end{equation*}
    
    The change points positions, $\tau_l$, are obtained recursively by assuming that $\tau_0 = 1$, and $\tau_l = m_l + \tau_{l-1} + w$ for $l = 1, \dotsc, k$.
    
    \item[(iii)] Distribution of the constant level $\alpha_l$: Given $\Tcp = k$, $\alpha_l$ is generated from the probability density function $\pi_0$ on $\mathbb{R}$ independently, where $\pi_0(\alpha_l) \propto 1$ for $\alpha_l \in \mathbb{R}$, and $l = 1, \dotsc, k+1$. 

    \item[(iv)] Finally, the constant-wise structure, $\bt_\ic(\ib)$, is then defined also based on the random quantities generated accordingly to their distribution defined in (i) - (iii). 

    \item[(v)] The baseline distribution $G_0$ is defined based on the distributions given in (i) - (iii): 
    \begin{align*}
    G_0(d\bt) &= \underbrace{P(\Tcp = k)}_{(i)} \underbrace{\left(\frac{\Gamma(m_0 + 1)}{\prod_{i=1}^{k} \Gamma(m_i + 1)} \left(\frac{1}{k+1}\right)^{m_0}\right)}_{(ii)} \underbrace{\prod_{l=1}^{k+1}\pi_0(\alpha_l)d\alpha_l}_{(iii)},\nonumber\\
    \end{align*}

\noindent where $dx$ represents an infinitesimal change in $x$. Therefore, $\pi_0(\alpha_l)d\alpha_l$ corresponds to the probability of observing the infinitesimal interval in the neighbourhood of $\alpha_l$.  

Because $\pi_0(\alpha_l) \propto 1$ for $l = 1, \dotsc, k + 1$, then $ G_0(d\bt) \propto P(\Tcp = k) \left(\frac{\Gamma(m_0 + 1)}{\prod_{i=1}^{k} \Gamma(m_i + 1)} \left(\frac{1}{k+1}\right)^{m_0}\right)$.

\end{itemize}

Note that, as mentioned, the distribution on the constant-wise structures, $G$, is discrete. Therefore, observations in cluster $\icluster,$ for $\icluster = 1, \dotsc, \Tcluster$ are assumed to share the same constant-wise function $\bt_{\icluster}$.
Parameter estimation for the model is achieved in a Bayesian framework via a Gibbs sampler.

\subsection{Bayesian Inference}
\label{sec:BayesianInference}

The vector with the observed data is denoted by $\boldsymbol{Y} =(\boldsymbol{Y}_{1}, \dotsc,\boldsymbol{Y}_{\Tc})^T$, where  $\boldsymbol{Y}_{\ic} = (Y_{\ic1}, \dotsc, Y_{\ic\Tb})^T$ for all observations $\ic = 1, \dotsc, \Tc$. $\underrightarrow{\bt} = (\bt_1, \dotsc, \bt_\Tc)$ is the collection of all constant-wise functions across all $\Tc$ observations. Let $\boldsymbol{\Tcp} = (\Tcp_1,\dotsc,\Tcp_\Tc)$ be the number of change points. We define the set of all change-point positions as $\boldsymbol{\tau} = (\boldsymbol{\tau}^{(\ic)}, \ic = 1, \dotsc, \Tc)$ with $\boldsymbol{\tau}^{(\ic)} = (\tau_1^{(\ic)}, \dotsc, \tau_{\Tcp_\ic}^{(\ic)})$, and $\boldsymbol{\alpha} = (\boldsymbol{\alpha}^{(\ic)}, \ic = 1, \dotsc, \Tc)$ as the set of all intercept parameters with $\boldsymbol{\alpha}^{(\ic)} = (\alpha_1^{(\ic)}, \dotsc, \alpha_{\Tcp_\ic + 1}^{(\ic)})^T$. Let $X_0^{(n)}$ be the $\Tb \times K_\ic + 1$ design matrix for $\boldsymbol{\alpha}^{(\ic)}$.

The prior distributions for the intercepts $\boldsymbol{\alpha}^{(\ic)}$ and Dirichlet Process hyperparameters, $\alpha_0$ and $\lambda$, are respectively $\pi_0$, $\pi_1$ and $\pi_2$ so that:

\begin{equation*}
    \pi_1(\alpha_0) = \mathrm{gamma}(a_{\alpha_0},b_{\alpha_0}),
\end{equation*}
and
\begin{equation*}
    \pi_2(\lambda) = \mathrm{gamma}(a_{\lambda},b_{\lambda}).
\end{equation*}
The prior distribution for the intercepts, $\pi_0$, is improper to provide analytical simplifications in the calculations for their posterior conditional distributions.

The variance components, $\boldsymbol{\sigma}^2 = (\sigma_1^2, \dotsc, \sigma_\Tc^2)$, are given independent inverse-gamma priors, such that

\begin{equation*}
    \pi_3(\boldsymbol{\sigma}^2) = \prod_{\ic = 1}^{\Tc}\mathrm{Inverse\;Gamma}(a_{\sigma^2},b_{\sigma^2}).
\end{equation*}

\subsubsection{Gibbs sampler}

In this section we present the updating steps for the estimation of the parameters $\bt_\ic,$ $\boldsymbol{\sigma}^2,$ $(\boldsymbol{\alpha}^{(\icluster)}, \Tcp_\icluster, \boldsymbol{\tau}^{(\icluster)}),$ $\lambda$ and $\alpha_0$ for $\ic = 1, \dotsc, \Tc$ and $\icluster = 1, \dotsc, \Tcluster$, where $\icluster$ denotes a individual cluster and $\Tcluster$ the total number of clusters. Each step involves calculating the full conditional distributions (see Appendix for derivation details).

\paragraph{Step 1: Update $\bt_\ic$}
The following expression demonstrates the clustering capability of the Dirichlet Process prior on the constant-wise structures $\bt_{\ic}$. The current value of  $\bt_{\ic}$ can be selected to be one of the existing $\bt_{\icluster}$ with positive probability $\sum_{j=1}^{\Tcluster}q_{\ic,j}/({q_{\ic,0} + \sum_{j=1}^{\Tcluster}q_{\ic,j}})$. In case in which observation $\ic$ does not belong to any existing clusters, a new $\bt_{\ic}$ is generate from the posterior distribution $G^{*}(\bt_\ic)$ as shown in Equation (\ref{eq:post_theta}).

The posterior of $\bt_\ic$ conditional on $\bt_{-\ic} = (\bt_{1}, \dotsc, \bt_{\ic - 1}, \bt_{\ic + 1}, \dotsc, \bt_{\Tc})$ is given by 

\begin{equation}
\label{eq:post_theta}
P(\bt_\ic \mid \bt_{-\ic}, \boldsymbol{Y}) = \frac{q_{\ic, 0}G^{*}(d\bt_\ic) + \sum_{j=1}^{\Tcluster}q_{\ic,j}\delta_{\bt_{(j)}}}{q_{\ic,0} + \sum_{j=1}^{\Tcluster}q_{\ic,j}},
\end{equation}
where 
\begin{equation*}
q_{\ic,0} = \int_{\boldsymbol{\Theta}}\ell(\boldsymbol{Y}_\ic \mid \bt_{\ic}) G_0(d\bt_\ic) \alpha_0/(\alpha_0 + \Tc - 1)
\ \text{and} \
q_{\ic,j} = \ell(\boldsymbol{Y}_\ic \mid \bt_{\ic}) \Tc_{\icluster}/(\alpha_0 + \Tc - 1),
\end{equation*}
define the mixing weights when observation $\ic$ forms a new cluster and when observation $\ic$ belongs to an existing cluster, respectively. Additionally, 
\begin{equation*}
   G^{*}(d\bt_\ic) = \frac{\ell(\boldsymbol{Y}_\ic \mid \bt_{\ic}) G_0(d\bt_\ic)}{\int_{\boldsymbol{\Theta}}\ell(\boldsymbol{Y}_\ic \mid \bt_{\ic}) G_0(d\bt_\ic)}
\end{equation*}
is the posterior of $\bt_\ic$ given that a new cluster is formed by observation $\ic$. Since $Y_{\ic\ib} \sim N(\alpha_l^{(\ic)}, \sigma^2_\ic)$, we have that $\ell(\boldsymbol{Y}_\ic \mid \bt_{\ic})$ represents the normal likelihood function corresponding to the observation $\boldsymbol{Y}_{\ic}$ after integrating out the variance component $\sigma^2_\ic$. 
Also, $\alpha_0$ corresponds to the precision hyperparameter for the Dirichlet Process and $\Tc_{\icluster}$ denotes the number of observations in cluster $\icluster$. The full expressions for $q_{\ic,0}$ and $q_{\ic,j}$ are given in detail in the Appendix Equations (\ref{eq:qn0}) and (\ref{eq:qnj}). 

\paragraph{Step 2: Update $\sigma^2_\ic$}
 Regardless of whether $\bt_\ic$ is a new value or an existing $\bt_{\icluster}$ (Step 1). The variance component for observation $\ic$ is updated using the full conditional of $\sigma^2_\ic$ given the other parameters.

\begin{align*}
P(\sigma^2_\ic \mid \bt_\ic, \boldsymbol{Y}_\ic) &\propto f(\boldsymbol{Y}_\ic \mid \bt_\ic, \sigma_\ic^2) \pi_3(\sigma^2_\ic) \\  \nonumber
&= \mathrm{Inverse\;Gamma}\left(\frac{\Tc}{2}+a_{\sigma^2},\frac{(\boldsymbol{Y}_\ic - X_0\boldsymbol{\alpha}^{(\ic)})^T(\boldsymbol{Y}_\ic - X_0\boldsymbol{\alpha}^{(\ic)})}{2} + \frac{1}{b_{\sigma^2}}\right).
\end{align*}

\paragraph{Step 3: Update  $(\Tcp_\icluster, \boldsymbol{\tau}^{(\icluster)}, \boldsymbol{\alpha}^{(\icluster)})$}
$\underrightarrow{\bt}$ uniquely determines the collection of parameters $(\boldsymbol{\Tcp}, \boldsymbol{\tau}, \boldsymbol{\alpha})$, and as mentioned, it contains several identical elements. Therefore, $(\boldsymbol{\Tcp}, \boldsymbol{\tau}, \boldsymbol{\alpha})$ also contains identical elements. In this step, we provide the updating procedures for the $\Tcluster$ distinct components of $(\boldsymbol{\Tcp}, \boldsymbol{\tau}, \boldsymbol{\alpha})$, defined by $(\Tcp_\icluster, \boldsymbol{\tau}^{(\icluster)}, \boldsymbol{\alpha}^{(\icluster)})$, for $\icluster = 1, \dotsc, \Tcluster$, where $\Tcluster$ is the number of clusters at the current update of the Gibbs sampler. Considering the hierarchical structure for the distributions of $\boldsymbol{\alpha}^{(\icluster)}$ and $\boldsymbol{\tau}^{(\icluster)}$, which both depends on the value of $\Tcp_\icluster$, we first update $\Tcp_\icluster$ from the posterior marginal probability function as follows:

\begin{equation*}
P(\Tcp_\icluster = k) \propto P(\Tcp = k) \sum_{(\ib_1, \dotsc, \ib_{k+1})} v(\ib_1, \dotsc, \ib_{k+1}), 
\end{equation*}
where
\begin{equation*}
v(\ib_1, \dotsc, \ib_{k+1}) = \exp{(\tilde{H}(\ib_1, \dotsc, \ib_{k+1}))}\frac{\Gamma(\ib_0 + 1)}{\prod_{l=1}^{k+1}\Gamma(\ib_l + 1)}\left(\frac{1}{k+1}\right)^{\ib_0}.
\end{equation*}

\noindent The full expression of $v(\ib_1, \dotsc, \ib_{k+1})$ is given in detail in the Appendix Equations (\ref{eq:v_def1}) and (\ref{eq:v_def2}).
\\

Then, we update $\boldsymbol{\tau}^{(\icluster)}$ given $\Tcp_{\icluster} = k$ using the probabilities $P(\ib_1, \dotsc, \ib_{k+1})$,
where
\begin{equation*}
P(\ib_1, \dotsc, \ib_{k+1}) \propto v(\ib_1, \dotsc, \ib_{k+1}).    
\end{equation*}

\noindent This is carried out by exhaustively listing all combinations and numerically computing the corresponding probabilities.
\\

Finally, $\boldsymbol{\alpha}^{(\icluster)}$ given $\boldsymbol{\tau}^{(\icluster)}$ and $\Tcp_{\icluster} = k$ is updated based on the full conditional distribution:

\begin{align*}
P(\boldsymbol{\alpha}^{(\icluster)} \mid \boldsymbol{\tau}^{(\icluster)}, \Tcp_\icluster, \boldsymbol{Y}_\icluster) &\propto f(\boldsymbol{Y}_\icluster \mid \boldsymbol{\alpha}^{(\icluster)}, \boldsymbol{\sigma}^2)  \pi(\boldsymbol{\alpha}^{(\icluster)}) \nonumber\\
&=\mathrm{Normal}\left(V^{-1}_\icluster X_{0,\icluster}^T\boldsymbol{Y}_\ic, V^{-1}_\icluster \sum_{\ic \in C_\icluster}\sigma^{-2}_\ic\right),
\end{align*}
where $\boldsymbol{Y}_\icluster$ represents the observations in cluster $\icluster$.

\paragraph{Step 4: Update $\lambda$}
The update of $\lambda$ is given by the following full conditional distribution and it is carried out by the Metropolis-Hasting algorithm; that is, we generate proposals from a gamma distribution and accept them with some probability based on an acceptance ratio.

\begin{align*}
P(\lambda \mid \Tcp, \boldsymbol{Y}) &\propto \pi(\lambda) \prod_{\ic = 1}^{\Tc}P(\Tcp_{\ic} = k) 
\nonumber \\
&= \frac{1}{b^a\Gamma(a)}\lambda^{a - 1}\exp\left\{\frac{\lambda}{b}\right\} \times \prod_{\ic = 1}^{\Tc} \frac{\lambda^k / k!}{\sum_{l = 0}^{k^*} \lambda^l / l!} \nonumber\\
&\propto \frac{\lambda^{a - 1 + \sum_{\ic = 1}^{\Tc}\Tcp_\ic }}{\left(\sum_{l = 0}^{k^*} \lambda^l / l!\right)^\Tc} \exp\left\{\frac{\lambda}{b}\right\},
\end{align*}
where $a$ and $b$ are the prior hyperparameters previously defined as $a_\lambda$ and $b_\lambda$.

\paragraph{Step 5: Update $\alpha_0$}
The update of $\alpha_0$ is carried out using the procedure described in \cite{Escobar1998}.

\begin{itemize}
    \item Sample $u \sim \mathrm{Beta}(\alpha_0^{t-1} + 1, \Tc)$
    \item Draw $\alpha_0$ from the mixture $\pi_u \times \mathrm{Gamma}(a + \Tcluster, (1/b - \log(u))^{-1}) + (1- \pi_u)\times \mathrm{Gamma}(a + \Tcluster - 1, (1/b - \log(u))^{-1})$,
\end{itemize}
where $a$ and $b$ are the prior hyperparameters previously described as $a_{\alpha_0}$ and $b_{\alpha_0}$, and $\Tcluster$ is the number of clusters at the current update of the Gibbs sampler and the probability membership is $$\pi_u = \frac{a + \Tcluster - 1}{\Tc(1/b - \log(u))}.$$

\section{Simulations}

We evaluate the performance of our method through three simulated scenarios. We varied one of the parameters for each simulated scenario while fixing the others, as shown in Table \ref{tab:scenarios}. 
We apply our method to the 96 randomly generated datasets based on the model in Equation (\ref{eq:model}) considering the initialization for the Gibbs sampler as described in Section \ref{sec:gibbs_initial}. Then, considering the evaluation metrics described in Section \ref{sec:performances}, we assess our method's performance and results are presented in the following Sections \ref{sec:scenario1}, \ref{sec:scenario2}, and \ref{sec:scenario3}.

\begin{table}[htbp]
\centering
\caption{Scenario configurations. We varied one of the parameters for each simulated scenario (shown by the * symbol), keeping the others fixed. The fixed parameters are $\Tc = 25$, $\Tb = 50$, $\Tcluster = 2$, and $\Tcp = 2$. The possible values for the number of data sequences ($\Tc$) are $10, 25, 50$, and for the number of locations ($\Tb$) are $50, 100, 200$. Additionally, while varying the number of data sequences, we considered one simulated scenario where variance components ($\sigma^2_\ic, \ic = 1, \dotsc, \Tc$) were generated from an inverse-gamma with a mean equal to 0.05, and another with a mean equal to 0.5.}
\label{tab:scenarios}
\begin{tabular}{lccc}  
\toprule
Scenario  & $\sigma^2$ (average) & $\Tc$ & $\Tb$ \\
\midrule
1       & 0.05  & *  & 50  \\
2       & 0.50   & *  & 50 \\
3       & 0.05  & 25 & *  \\
\bottomrule
\end{tabular}
\end{table}

\subsection{Gibbs sampler initialization and implementation}
\label{sec:gibbs_initial}

This section describes the initialization for the Gibbs sampler and some details about our algorithm's implementation. For the simulation scenarios and real data analysis, the hyperparameters for the prior distribution of the variance components were specified as $a_{\sigma^2} = 2$ and $b_{\sigma^2} = 1000$. For the prior distributions of $\lambda$ and $\alpha_0$, the hyperparameters were $a_\lambda = a_{\alpha_0} = 2$ and $b_\lambda = b_{\alpha_0} = 1000$. The minimum number of locations in each segment between change points, $w$, was set to 10 in Scenario 1, and was set to 10, 20, and 50 for Scenario 2, when considering $M = 50, 100,$ and $200$ respectively.

To enable convergence diagnosis for the Gibbs sampler we employed two chains with different initial values for each simulated scenario. The first chain starts from the true settings, that is, we consider the parameter values used to generate the datasets as initial values for our algorithm, whereas for the second chain we initialize the 
Gibbs sampler from the true parameter values plus a small perturbation. For instance, the initial values for the intercepts of each cluster are initialized from the true setting plus 1.5. The position of the change points for each cluster starts from two points above the ground truth and the variance components are initialized using generated values from an inverse-gamma distribution with twice the average used to generate the true variance components.

The simulations and computations for the Gibbs sampler algorithm were performed using Sharcnet's Graham cluster, with a single node consisting of two Intel E5-2683 v4 “Broadwell” with 2.1GHz processor base frequency, for an overall of 32 computing cores. The number of simulated datasets, S = 96, was chosen as a multiple of the number of cores. The computations were performed on [CentOS 7], with R version [4.2.1 “Funny-Looking Kid”] \citep{R2022}, using the parallel package \citep{R2022} to simulate and to compute the Gibbs sampler for independent datasets simultaneously, the extraDistr package \citep{extraDistr2022} to generate samples from inverse-gamma distributions, the RcppAlgos package \citep{RcppAlgos2022} to generate all possible partitions for the number of points in each segment between two change points, the MASS package \citep{MASS2022} to generate samples from multivariate normal distributions, and the package FDRSeg \citep{Rosenberg2007} to calculate the V-measure. It is worth mentioning that our algorithm is implemented as the R package BayesCPclust and is available at
\texttt{https://CRAN.R-project.org/package=BayesCPclust}.

\subsection{Performance metrics}
\label{sec:performances}

For each chain, simulated setting, and randomly generated dataset, we employ our method with 5000 iterations to estimate change points and perform clustering. We consider a burn-in of $50\%$ of the size of the chains, and we thin our remaining samples by keeping only every 25th iteration. This procedure ensures that our samples are not highly correlated. For the 200 remaining samples, we calculate the posterior mean for each parameter, except for the discrete variables, such as cluster assignments, number of clusters, number of change points, and their locations, where we choose the most frequent value, the posterior mode.

To evaluate our method's performance concerning intercept estimation, we compute the average of the posterior means for each intercept and compare its value to the true settings considered when generating the datasets. We report the posterior means standard error and the average interval length of $95\%$ equal-tailed credible intervals taken over the 96 datasets to measure the precision of our estimates. Additionally, we report the MAD (Mean Absolute Deviation) for the variance components estimates.

For the discrete variables, we report the proportion of datasets in which we correctly estimated the parameters. To evaluate the clustering performance of our proposed approach, we consider the V-measure \citep{Rosenberg2007}, which assesses observation-to-cluster assignments and measures the homogeneity and completeness of a clustering result. The V-measure ranges from zero to one, where results closer to one are considered adequate.

\subsection{Scenario 1: Varying the number of data sequences with $\sigma_\ic^2 \approx 0.05$}
\label{sec:scenario1}

\begin{figure}[htbp]
    \centering
    \includegraphics[width = \textwidth]{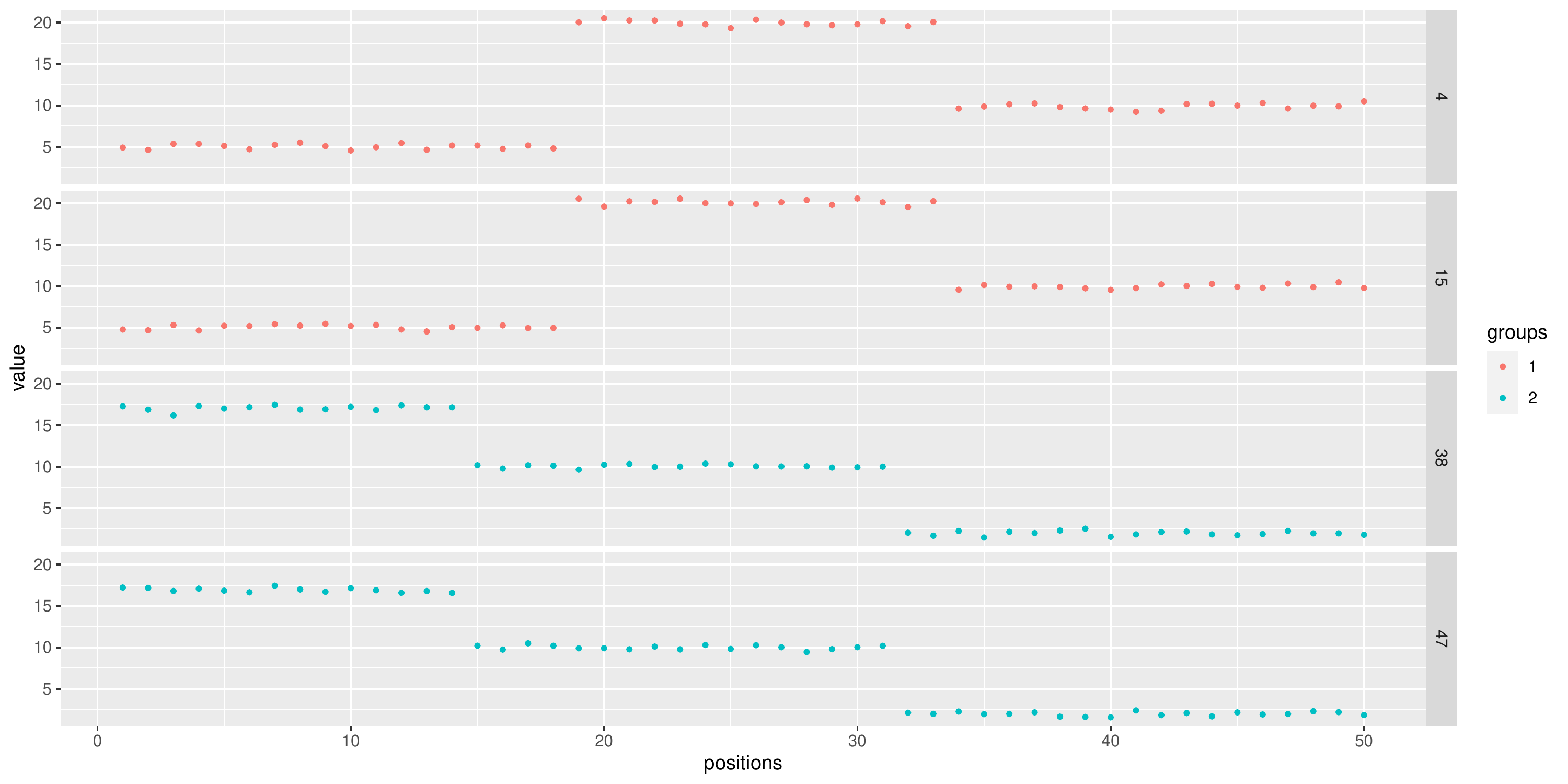}
    \caption{Simulation scenario 1. Data structure for four data sequences from one of the 96 randomly generated synthetic datasets for Scenario 1 ($\Tc = 50 \text{ and } \Tb = 50$) with $\sigma^2_n \approx 0.05$. Each panel presents the observed values for one data sequence. Observations from Cluster 1 are colored in red, whereas observations from Cluster 2 are colored in blue.}
    \label{fig:sec1_data}
\end{figure}

Figure \ref{fig:sec1_data} shows the data structure of four data sequences from one of the 96 randomly generated synthetic datasets for Scenario 1 when $N = 50$. In this scenario, we vary the number of data sequences considering $\Tc = 10, 25$, and $50$ while keeping the other parameters fixed as described in Table \ref{tab:scenarios}. Each panel represents one observation colored by their cluster assignment. Both clusters have two change points. Change-point locations for Cluster 1 are 19 and 34; for the second cluster, they are 15 and 32. Each segment between change points is defined by a constant level (5, 20, and 10) for the first cluster and (17, 10, and 2) for the second cluster.

\begin{figure}[ht]
    \centering
    \begin{subfigure}[b]{\textwidth}
        \centering
        \includegraphics[page = 1]{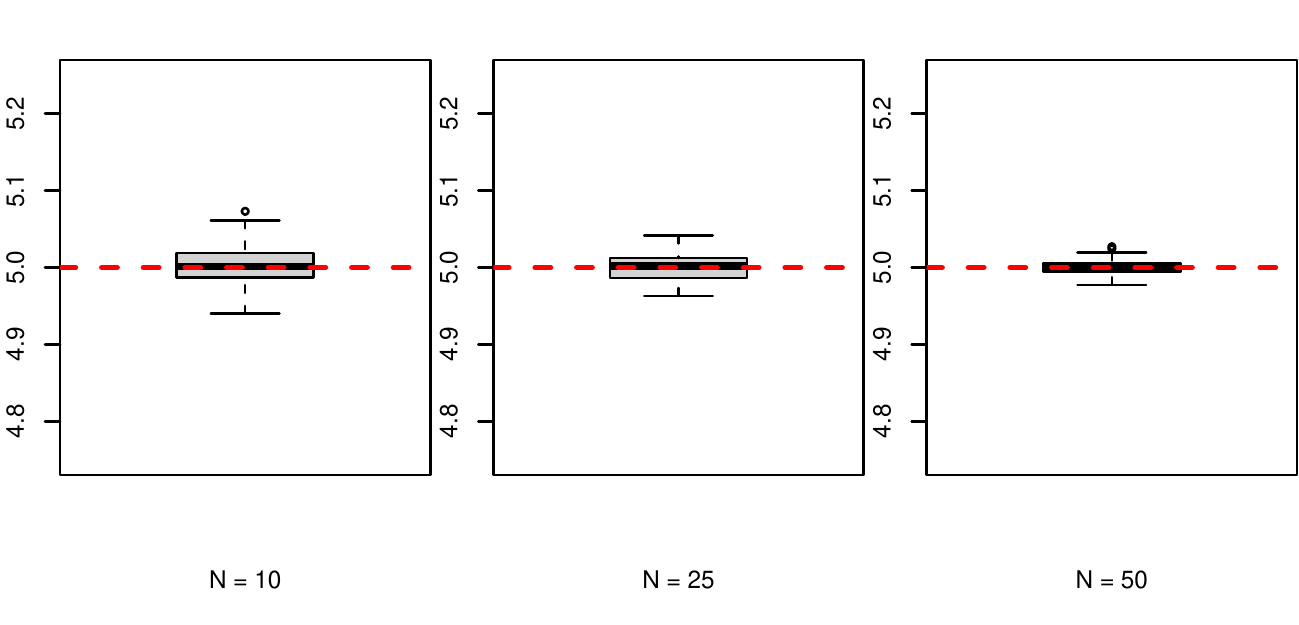}
        \caption{}
    \end{subfigure}
    \begin{subfigure}[b]{\textwidth}
        \centering
        \includegraphics[page = 2]{output_plots_scenario1_cluster1.pdf}
        \caption{}
    \end{subfigure}
    \begin{subfigure}[b]{\textwidth}
        \centering
        \includegraphics[page = 3]{output_plots_scenario1_cluster1.pdf}
        \caption{}
    \end{subfigure}
    \caption{Simulation scenario 1, Cluster 1. Boxplots of the posterior mean estimates for the intercepts (a) $\alpha_1$, (b) $\alpha_2$, and (c) $\alpha_3$ of our change-point model for Cluster 1 based on the 96 randomly generated synthetic data sets, when we vary the number of observations $\Tc$  with $\sigma^2_n \approx 0.05$. The red dashed lines correspond to the true parameter values for each intercept.}
    \label{fig:sc1_intercepts_cluster1}
\end{figure}

\begin{figure}[ht]
    \centering
    \begin{subfigure}[b]{\textwidth}
        \centering
        \includegraphics[page = 1]{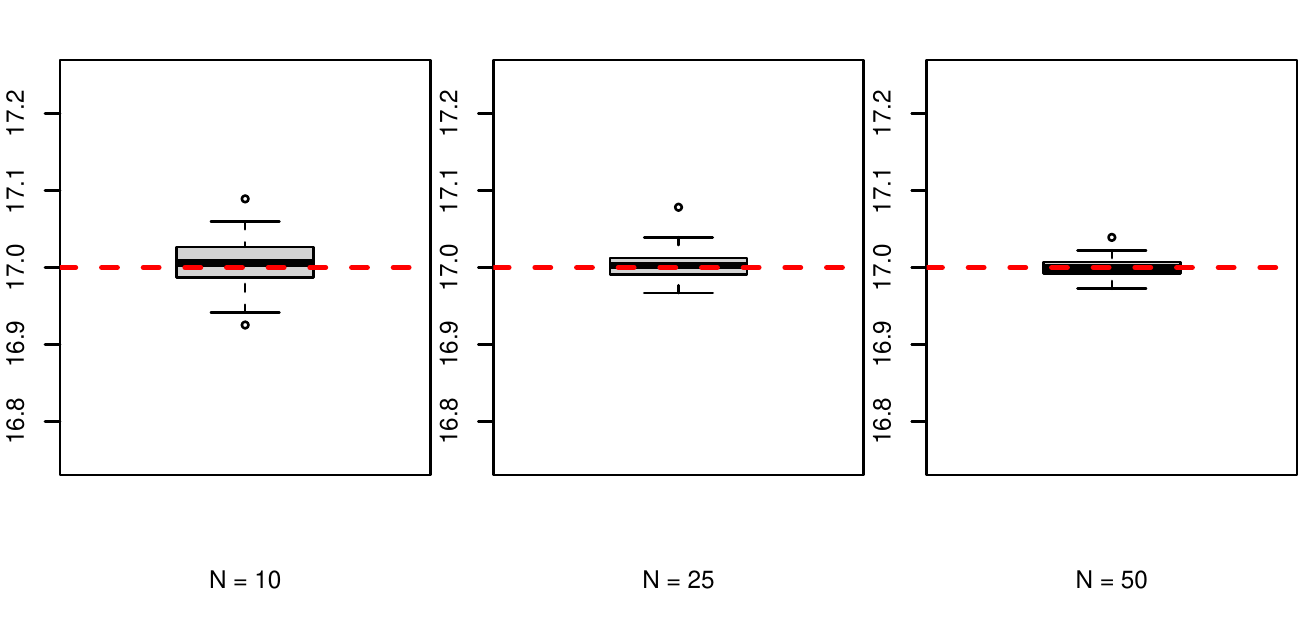}
        \caption{}
    \end{subfigure}
    \begin{subfigure}[b]{\textwidth}
        \centering
        \includegraphics[page = 2]{output_plots_scenario1_cluster2.pdf}
        \caption{}
    \end{subfigure}
    \begin{subfigure}[b]{\textwidth}
        \centering
        \includegraphics[page = 3]{output_plots_scenario1_cluster2.pdf}
        \caption{}
    \end{subfigure}
    \caption{Simulation scenario 1, Cluster 2. Boxplots of the posterior mean estimates for the intercepts (a) $\alpha_1$, (b) $\alpha_2$, and (c) $\alpha_3$ of our change-point model for Cluster 2 based on the 96 randomly generated synthetic data sets, when we vary the number of observations $\Tc$ with $\sigma^2_n \approx 0.05$ . The red dashed lines correspond to the true parameter values for each intercept.}
    \label{fig:sc1_intercepts_cluster2}
\end{figure}

\begin{table}[ht]
\centering
\caption{Simulation scenario 1. Posterior estimates for the intercepts of each cluster taken over 96 randomly generated synthetic datasets when we vary the number of observations ($\Tc$). The variance components in this scenario were sampled from an inverse gamma with small average (0.05). We report the average posterior mean estimates (Average) and standard error (SE) for each intercept, and we present the average interval length of $95\%$ credible intervals (Average CI size) taken over the 96 datasets.} 
\label{tab:sc1_itercepts}
\begin{tabular}{cllrrr}
  \toprule
Cluster & Parameter & $\Tc$ & Average & SE & Average CI size \\ 
  \midrule
 &  & 10 & 4.9997 & 0.0260 & 0.0874 \\ 
   & $\alpha_{1} = 5$ & 25 & 4.9997 & 0.0169 & 0.0597 \\ 
   &  & 50 & 5.0001 & 0.0101 & 0.0389 \\ 
   \cmidrule(r){2-6}
   &  & 10 & 20.0027 & 0.0285 & 0.0943 \\ 
  1 & $\alpha_{2} = 20$ & 25 & 20.0009 & 0.0217 & 0.0665 \\ 
   &  & 50 & 20.0010 & 0.0129 & 0.0425 \\ 
   \cmidrule(r){2-6}
   &  & 10 & 9.9977 & 0.0259 & 0.0897 \\ 
   & $\alpha_{3} = 10$ & 25 & 10.0000 & 0.0144 & 0.0628 \\ 
   &  & 50 & 10.0005 & 0.0118 & 0.0403 \\ 
   \midrule
   &  & 10 & 17.0069 & 0.0279 & 0.1031 \\ 
   & $\alpha_{1} = 17$ & 25 & 17.0027 & 0.0179 & 0.0658 \\ 
   &  & 50 & 17.0002 & 0.0113 & 0.0432 \\ 
   \cmidrule(r){2-6}
   &  & 10 & 10.0005 & 0.0298 & 0.0940 \\ 
  2 & $\alpha_{2} = 10$ & 25 & 9.9991 & 0.0159 & 0.0596 \\ 
   &  & 50 & 9.9988 & 0.0116 & 0.0387 \\ 
    \cmidrule(r){2-6}
   &  & 10 & 1.9994 & 0.0253 & 0.0876 \\ 
   & $\alpha_{3} = 2$ & 25 & 2.0003 & 0.0150 & 0.0568 \\ 
   &  & 50 & 2.0000 & 0.0095 & 0.0376 \\ 
   \bottomrule
\end{tabular}
\end{table}
Based on the methodology of \cite{Gelman1992}, the convergence of the chains for all parameters, after the burn-in period and thinning procedure was confirmed. Table \ref{tab:sc1_itercepts} and Figures \ref{fig:sc1_intercepts_cluster1} and \ref{fig:sc1_intercepts_cluster2} display the results for the posterior estimates for the intercepts of each cluster when the number of data sequences is $\Tc =$  10, 25, and 50, and the variance components were generated around 0.05. In this setting, our estimates were close to the true parameter values, showing that our proposed method retrieved the correct intercepts for each cluster. As the number of data sequences increases, the standard error of our intercept estimates and the average length of the $95\%$ credible intervals decreases, as expected. Considering that each data sequence has its own variance component and $\Tb$ is fixed, increasing the number of data sequences does not considerably improve the estimation of the variance components as shown in Table \ref{tab:sc1_MAD} by the mean absolute deviation.

\begin{table}[ht]
\centering
\caption{Simulation scenario 1. Mean absolute deviation (MAD) for the estimated variances (as the posterior means) when we vary the number of data sequences and the variance components were sampled from an inverse-gamma with average 0.05.} 
\label{tab:sc1_MAD}
\begin{tabular}{lr}
  \toprule
$\Tc$ & MAD \\ 
  \midrule
10 & 0.0073 \\ 
25 & 0.0091 \\ 
50 & 0.0079 \\ 
  \bottomrule
\end{tabular}
\end{table}

The change point locations associated with the two clusters were correctly estimated for all 96 datasets. The number of clusters, and the cluster assignment for each data sequence were correctly estimated for all 96 datasets resulting in all V-measures to be equal to one.

\subsection{Scenario 2: Varying the number of data sequences with $\sigma_\ic^2 \approx 0.50$}
\label{sec:scenario2}

\begin{figure}[htbp]
    \centering
    \includegraphics[width = \textwidth]{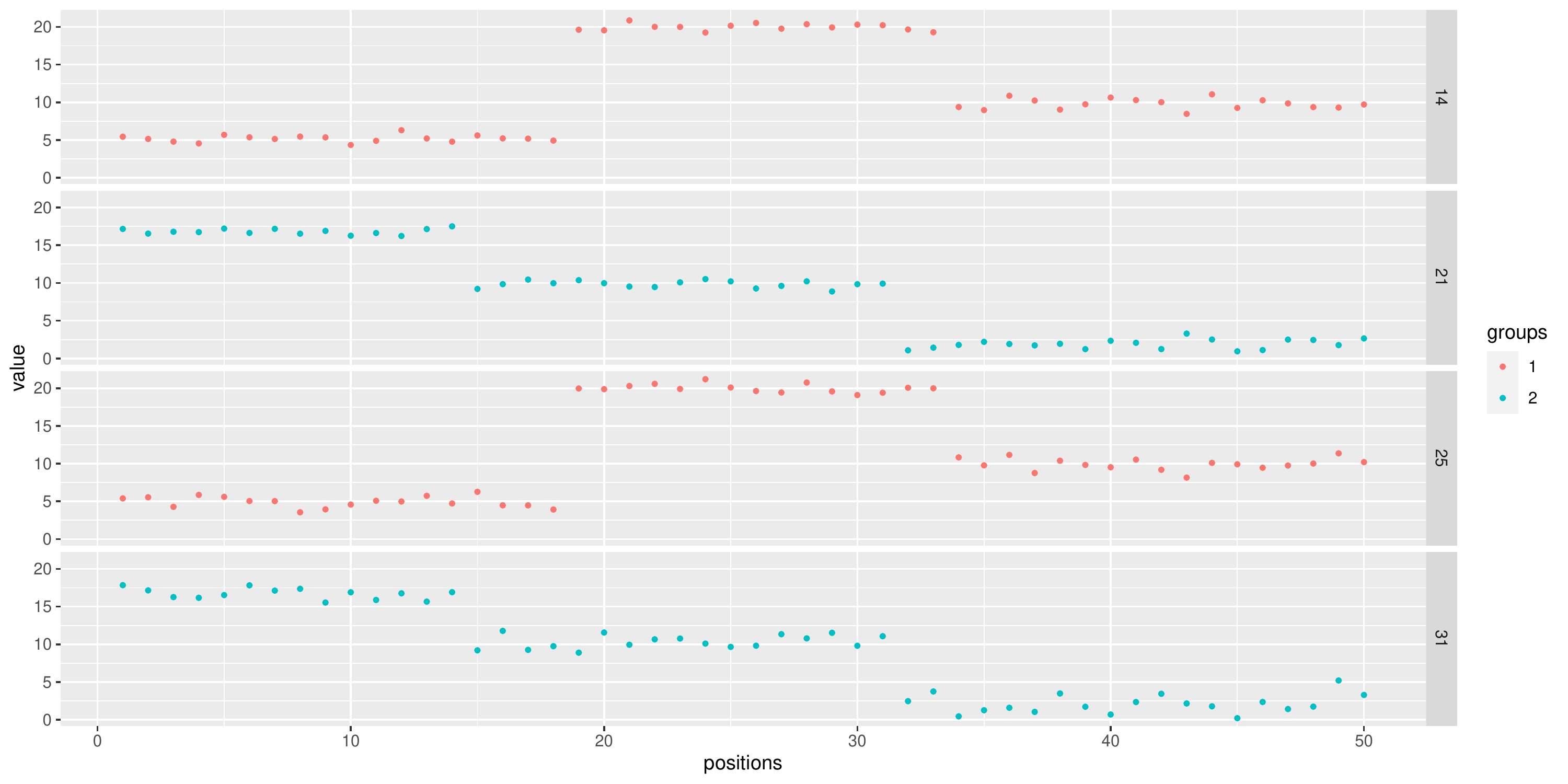}
    \caption{Simulation scenario 2. Data structure for four data sequences from one of the 96 randomly generated synthetic datasets for Scenario 2 ($\Tc = 50 \text{ and } \Tb = 50$) with $\sigma^2_n \approx 0.5$. Each panel presents the observed values for one data sequence. Observations from Cluster 1 are colored in red, whereas observations from Cluster 2 are colored in blue.}
    \label{fig:sc2_data}
\end{figure}

This section evaluates our method's performance with a higher data dispersion than in the previous section. We generate 96 datasets as in the last experiment for each possible value of $\Tc$; however, for this scenario, we sample the variance components from an inverse gamma with an average 10 times higher than in the simulation Scenario 1 as shown in Figure \ref{fig:sc2_data}. 

\begin{figure}[ht]
    \centering
    \begin{subfigure}[b]{\textwidth}
        \centering
        \includegraphics[page = 1]{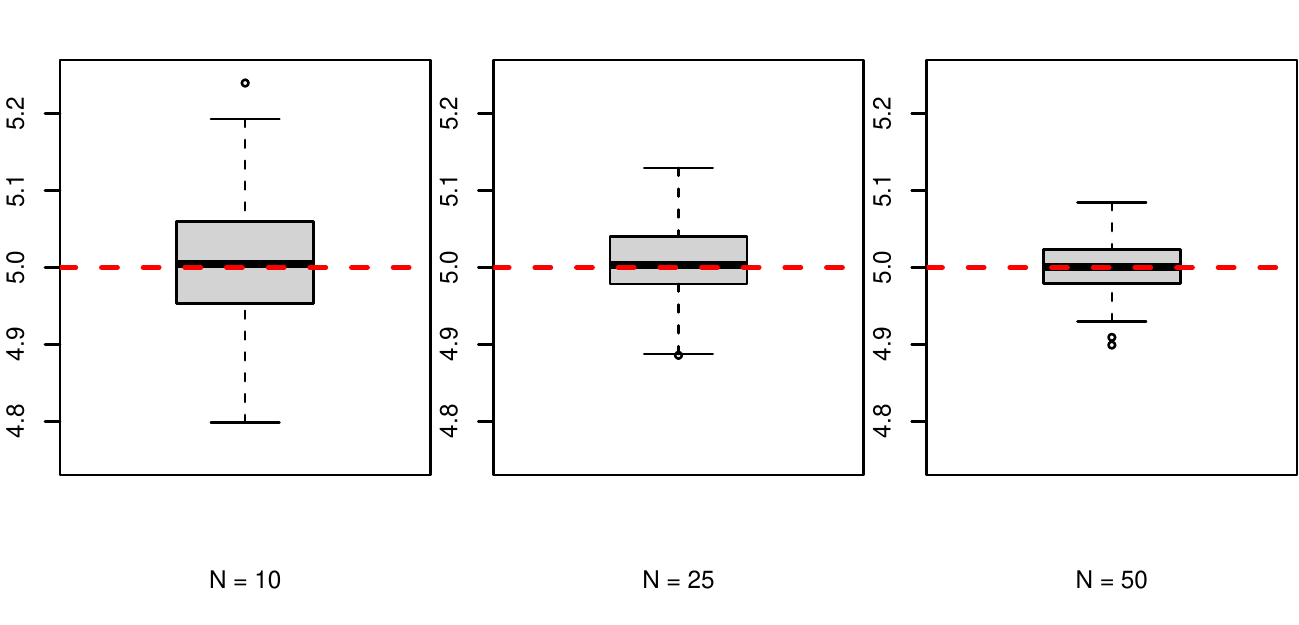}
        \caption{}
    \end{subfigure}
    \begin{subfigure}[b]{\textwidth}
        \centering
        \includegraphics[page = 2]{output_plots_scenario2_cluster1.pdf}
        \caption{}
    \end{subfigure}
    \begin{subfigure}[b]{\textwidth}
        \centering
        \includegraphics[page = 3]{output_plots_scenario2_cluster1.pdf}
        \caption{}
    \end{subfigure}
    \caption{Simulation scenario 2, Cluster 1. Boxplots of the posterior mean estimates for the intercepts (a) $\alpha_1$, (b) $\alpha_2$, and (c) $\alpha_3$ of our change-point model for Cluster 1 based on the 96 randomly generated synthetic data sets, when we vary the number of observations $\Tc$  with $\sigma^2_n \approx 0.5$. The red dashed lines correspond to the true parameter values for each intercept.}
    \label{fig:sc2_intercepts_cluster1}
\end{figure}

\begin{figure}[ht]
    \centering
    \begin{subfigure}[b]{\textwidth}
        \centering
        \includegraphics[page = 1]{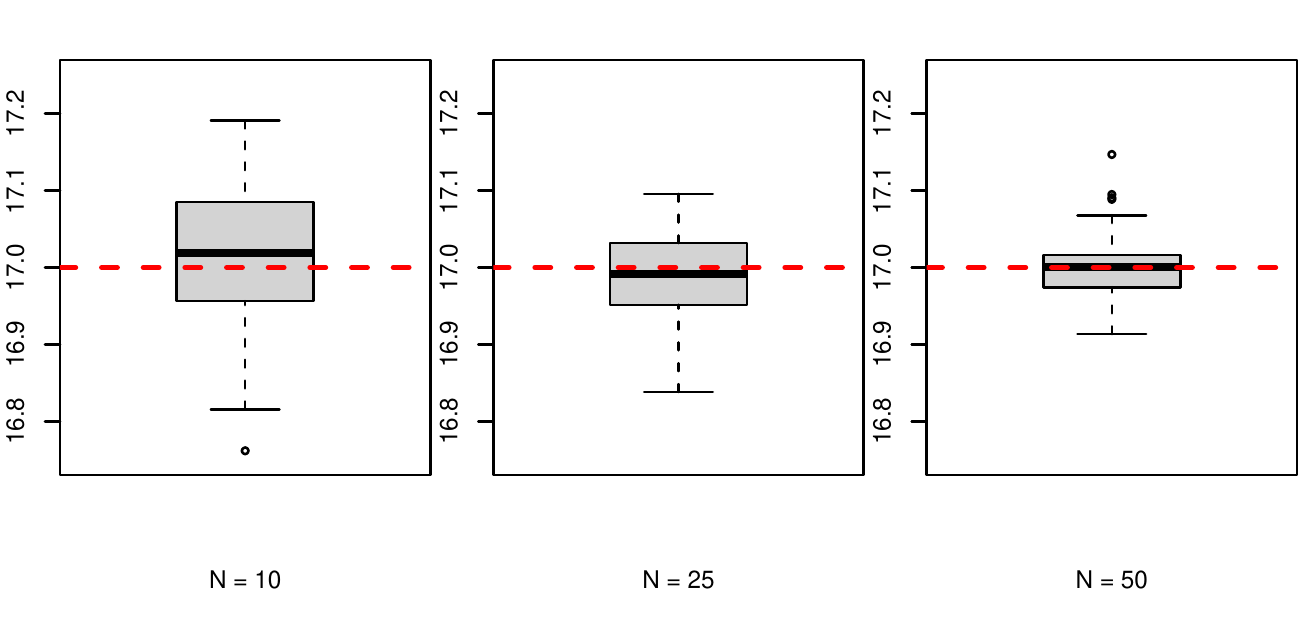}
        \caption{}
    \end{subfigure}
    \begin{subfigure}[b]{\textwidth}
        \centering
        \includegraphics[page = 2]{output_plots_scenario2_cluster2.pdf}
        \caption{}
    \end{subfigure}
    \begin{subfigure}[b]{\textwidth}
        \centering
        \includegraphics[page = 3]{output_plots_scenario2_cluster2.pdf}
        \caption{}
    \end{subfigure}
    \caption{Simulation scenario 2, Cluster 2. Boxplots of the posterior mean estimates for the intercepts (a) $\alpha_1$, (b) $\alpha_2$, and (c) $\alpha_3$ of our change-point model for Cluster 2 based on the 96 randomly generated synthetic data sets, when we vary the number of observations $\Tc$ with $\sigma^2_n \approx 0.5$ . The red dashed lines correspond to the true parameter values for each intercept.}
    \label{fig:sc2_intercepts_cluster2}
\end{figure}

It is worth mentioning that the convergence of the chains for all parameters in Scenario 2 was also confirmed using the methodology of \cite{Gelman1992}. Table \ref{tab:sc2_intercepts} and Figures \ref{fig:sc2_intercepts_cluster1} and \ref{fig:sc2_intercepts_cluster2} show the results for the posterior estimates for the intercepts of each segment between change points for the two clusters when the number of data sequences is $\Tc = 10$, $25$ and $50$ and the variance components were generated around 0.5. We observe that for each possible value for the number of data sequences, our approach correctly estimated the intercepts. However, the standard errors of our estimates are noticeably higher than in the previous scenario due to the increase in the data dispersion. Nonetheless, our method showed good performance, not only in estimating the intercepts for each cluster but also in correctly estimating the number of change points and their corresponding locations. Additionally, our method always recovered the true clustering configuration in our data, with all V-measures equal to one.

Furthermore, the mean absolute deviation for the variance components estimates remained stable, as in the previous scenario, suggesting that increasing the number of data sequences does not noticeably improve the precision of the variance component estimates as reported in Table \ref{tab:sc2_MAD}.

\begin{table}[ht]
\centering
\caption{Simulation scenario 2. Posterior estimates for the intercepts of each cluster taken over 96 randomly generated synthetic datasets when we vary the number of observations ($\Tc$). The variance components in this scenario were sampled from an inverse-gamma with average equals to 0.50. We report the average posterior mean estimates (Average) and standard error (SE) for each intercept, and we present the average interval length of $95\%$ credible intervals (Average CI size) taken over the 96 datasets.} 
\label{tab:sc2_intercepts}
\begin{tabular}{cllrrr}
  \toprule
Cluster & Parameter & $\Tc$ & Average & SE & Average CI size \\ 
  \midrule
 &  & 10 & 4.9990 & 0.0825 & 0.2773 \\ 
   & $\alpha_{1} = 5$ & 25 & 5.0086 & 0.0508 & 0.1801 \\ 
   &  & 50 & 5.0021 & 0.0352 & 0.1265 \\ 
      \cmidrule(r){2-6}
   &  & 10 & 20.0077 & 0.0906 & 0.3015 \\ 
  1 & $\alpha_{2} = 20$ & 25 & 19.9953 & 0.0558 & 0.1992 \\ 
   &  & 50 & 19.9990 & 0.0481 & 0.1389 \\ 
      \cmidrule(r){2-6}
   &  & 10 & 9.9929 & 0.0831 & 0.2821 \\ 
   & $\alpha_{3} = 10$ & 25 & 10.0056 & 0.0493 & 0.1849 \\ 
   &  & 50 & 9.9956 & 0.0374 & 0.1315 \\ 
        \midrule
   &  & 10 & 17.0218 & 0.0882 & 0.3246 \\ 
   & $\alpha_{1} = 17$ & 25 & 16.9916 & 0.0526 & 0.2000 \\ 
   &  & 50 & 16.9989 & 0.0399 & 0.1431 \\ 
         \cmidrule(r){2-6}
   &  & 10 & 10.0021 & 0.0938 & 0.2962 \\ 
  2 & $\alpha_{2} = 10 $ & 25 & 10.0083 & 0.0586 & 0.1799 \\ 
   &  & 50 & 9.9958 & 0.0363 & 0.1273 \\ 
         \cmidrule(r){2-6}
   &  & 10 & 1.9977 & 0.0804 & 0.2784 \\ 
   & $\alpha_{3} = 2$ & 25 & 1.9946 & 0.0441 & 0.1737 \\ 
   &  & 50 & 2.0045 & 0.0317 & 0.1204 \\ 
   \bottomrule
\end{tabular}
\end{table}

\begin{table}[ht]
\centering
\caption{Simulation scenario 2. Mean absolute deviation (MAD) for the estimated variances (as the posterior means) when we vary the number of data sequences and the variance components were sampled from an inverse-gamma with average 0.5.} 
\label{tab:sc2_MAD}
\begin{tabular}{lr}
  \toprule
$\Tc$ & MAD \\ 
  \midrule
10 & 0.0736 \\ 
25 & 0.0784 \\ 
50 & 0.0833 \\ 
   \bottomrule
\end{tabular}
\end{table}

\subsection{Scenario 3: Varying the number of locations} 
\label{sec:scenario3}

\begin{figure}[ht]
    \centering
    \begin{subfigure}[b]{\textwidth}
        \centering
        \includegraphics[page = 1]{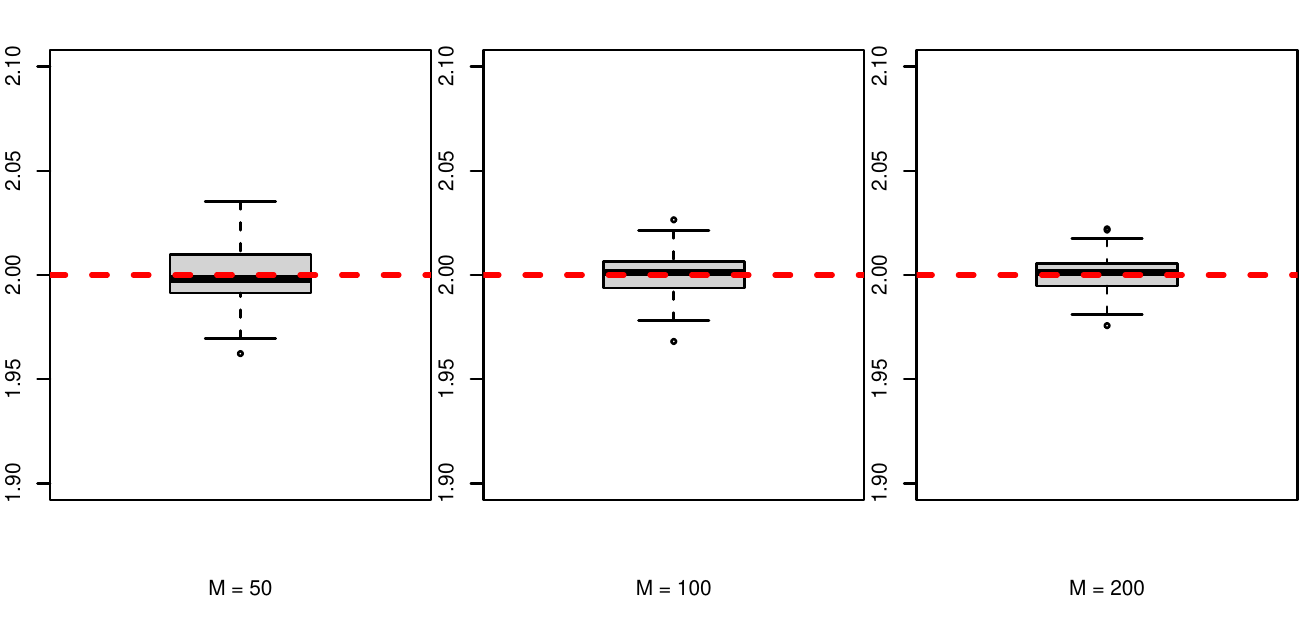}
        \caption{}
    \end{subfigure}
    \begin{subfigure}[b]{\textwidth}
        \centering
        \includegraphics[page = 2]{output_plots_scenario3_cluster1.pdf}
        \caption{}
    \end{subfigure}
    \begin{subfigure}[b]{\textwidth}
        \centering
        \includegraphics[page = 3]{output_plots_scenario3_cluster1.pdf}
        \caption{}
    \end{subfigure}
    \caption{Simulation scenario 3, Cluster 1. Boxplots of the posterior mean estimates for the intercepts (a) $\alpha_1$, (b) $\alpha_2$, and (c) $\alpha_3$ of our change-point model for Cluster 2 based on the 96 randomly generated synthetic data sets, when we vary the number of locations $\Tb$. The red dashed lines correspond to the true parameter values for each intercept.}
    \label{fig:sc3_intercepts_cluster1}
\end{figure}

\begin{figure}[ht]
    \centering
    \begin{subfigure}[b]{\textwidth}
        \centering
        \includegraphics[page = 1]{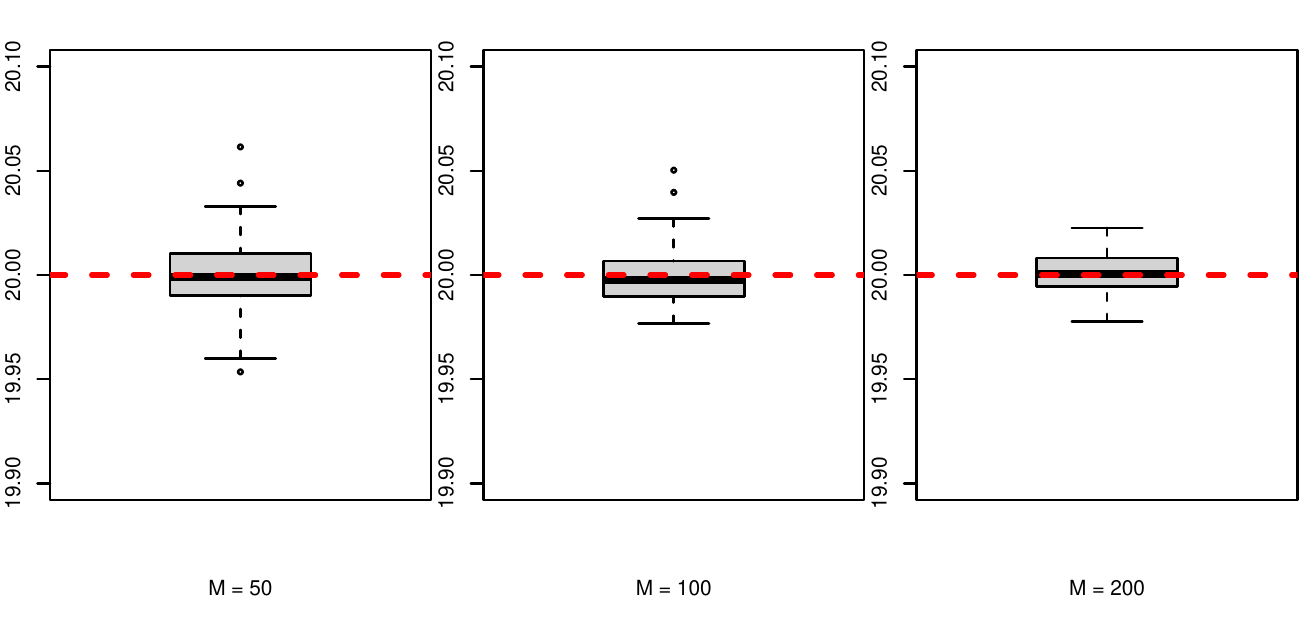}
        \caption{}
    \end{subfigure}
    \begin{subfigure}[b]{\textwidth}
        \centering
        \includegraphics[page = 2]{output_plots_scenario3_cluster2.pdf}
        \caption{}
    \end{subfigure}
    \begin{subfigure}[b]{\textwidth}
        \centering
        \includegraphics[page = 3]{output_plots_scenario3_cluster2.pdf}
        \caption{}
    \end{subfigure}
    \caption{Simulation scenario 3, Cluster 2. Boxplots of the posterior mean estimates for the intercepts (a) $\alpha_1$, (b) $\alpha_2$, and (c) $\alpha_3$ of our change-point model for Cluster 2 based on the 96 randomly generated synthetic data sets, when we vary the number of locations $\Tb$. The red dashed lines correspond to the true parameter values for each intercept.}
    \label{fig:sc3_intercepts_cluster2}
\end{figure}

In this Section, we present the performance results of our method when the number of locations is $\Tb = $ 50, 100, and 200. Table \ref{tab:sc3_intercepts} and Figures \ref{fig:sc3_intercepts_cluster1} and \ref{fig:sc3_intercepts_cluster2} present the results for the intercepts for each case in the Scenario 3. As in the previous scenarios, convergence of the chains for all parameters was confirmed. Based on the results, our approach correctly estimated the intercepts for each cluster and showed that as the number of locations increased, the precision of our estimates improved. Once again, our method correctly estimated the number of change points and change-point positions for all generated datasets. In addition, all V-measures were equal to one, showing that our model recovered the true clustering configuration in our data.

Furthermore, in this scenario, we observed an increase in the precision of our estimates for the variance components, as shown in Table \ref{tab:sc3_MAD} as $\Tb$ increases. As discussed in the previous scenarios, the number of data sequences minimally affects the precision of our variance estimates since each data sequence has its variance component. However, by increasing the number of locations, we noted a decrease in the mean absolute deviation for our estimates, suggesting that the number of locations considerably affects the estimation of the variance components.

\begin{table}[ht]
\centering
\caption{Simulation scenario 3. Posterior estimates for the intercepts of each cluster taken over 96 randomly generated synthetic datasets when we vary the number of locations ($\Tb$). We report the average posterior mean estimates (Average) and standard error (SE) for each intercept, and we present the average interval length of $95\%$ credible intervals (Average CI size) taken over the 96 datasets.} 
\label{tab:sc3_intercepts}
\begin{tabular}{cllrrr}
  \toprule
Cluster & Parameter & $\Tb$ & Average & SE  & Average CI size \\ 
  \midrule
 &  & 50 & 1.9999 & 0.0152 & 0.0574 \\ 
   & $\alpha_{1} = 2$ & 100 & 2.0003 & 0.0099 & 0.0416 \\ 
   &  & 200 & 2.0004 & 0.0085 & 0.0323 \\ 
      \cmidrule(r){2-6}
   &  & 50 & 14.9996 & 0.0165 & 0.0560 \\ 
  1 & $\alpha_{2} = 15$ & 100 & 15.0014 & 0.0103 & 0.0412 \\ 
   &  & 200 & 15.0011 & 0.0083 & 0.0319 \\ 
      \cmidrule(r){2-6}
   &  & 50 & 6.9966 & 0.0157 & 0.0619 \\ 
   & $\alpha_{3} = 7$ & 100 & 7.0013 & 0.0138 & 0.0480 \\ 
   &  & 200 & 7.0012 & 0.0068 & 0.0271 \\ 
     \midrule
   &  & 50 & 19.9997 & 0.0181 & 0.0615 \\ 
   & $\alpha_{1} = 20$ & 100 & 19.9990 & 0.0126 & 0.0455 \\ 
   &  & 200 & 20.0009 & 0.0100 & 0.0340 \\ 
      \cmidrule(r){2-6} 
   &  & 50 & 5.0008 & 0.0147 & 0.0560 \\ 
  2 & $\alpha_{2} = 5$ & 100 & 5.0005 & 0.0119 & 0.0447 \\ 
   &  & 200 & 4.9992 & 0.0067 & 0.0266 \\ 
      \cmidrule(r){2-6}
   &  & 50 & 12.0014 & 0.0122 & 0.0526 \\ 
   & $\alpha_{3} = 12$ & 100 & 11.9990 & 0.0108 & 0.0368 \\ 
   &  & 200 & 11.9995 & 0.0080 & 0.0341 \\ 
   \bottomrule
\end{tabular}
\end{table}

\begin{table}[ht]
\centering
\caption{Simulation scenario 3. Mean absolute deviation (MAD) for the estimated variances (as the posterior means) when we vary the number of locations.} 
\label{tab:sc3_MAD}
\begin{tabular}{lr}
  \toprule
$\Tb$ & MAD \\ 
  \midrule
50 & 0.0076 \\ 
100 & 0.0057 \\ 
200 & 0.0041 \\ 
   \bottomrule
\end{tabular}
\end{table}

\clearpage

\section{Real data analysis}
\label{sec:real_data}

We further assessed the performance of our method in a real dataset. We apply our approach to a subset of the copy number genomic data analyzed in \cite{Leung2017} focusing on patient CRC2. The dataset consists of copy-number information for 45 cells (data sequences) from frozen primary tumour and liver metastases of colorectal cancer. Each data point in the dataset corresponds to the $\log_2$ ratio of reads aligned per 200-kb genomic bin per cell after GC correction. The $\log_2$ ratios provide an indication of the number of copies in each genomic bin. A $\log_2$ ratio greater than one means an amplification in the corresponding region. Genomic copy number alterations are common in many diseases including cancer, where deletions or amplifications of DNA segments can contribute to alterations in the expression of tumour-suppressor genes \citep{zhao2016, shao2019}. Identifying the number and locations of these alterations is essential for understanding cancer progression. As tumours evolve, differences in genomic profiles, including copy number, are expected between primary tumour and metastatic tumours \citep{nowell1976, ding2012, bambury2015, eirew2015, kridel2016}.

In this paper, for computational feasibility purposes, we focus our analysis on chromosomes 19, 20, and 21, corresponding to 583 genomic bins (locations), since it is a region with visible change points as observed in \cite{Leung2017}. The raw data (FASTQ files) are available publicly at NCBI Sequence Read Archive (SRA) under accession number SRP074289. Processed $\log_2$ ratios were kindly provided by the authors of \cite{Leung2017} upon our request.

\begin{figure}[htbp]
    \centering
    \includegraphics[width = \textwidth]{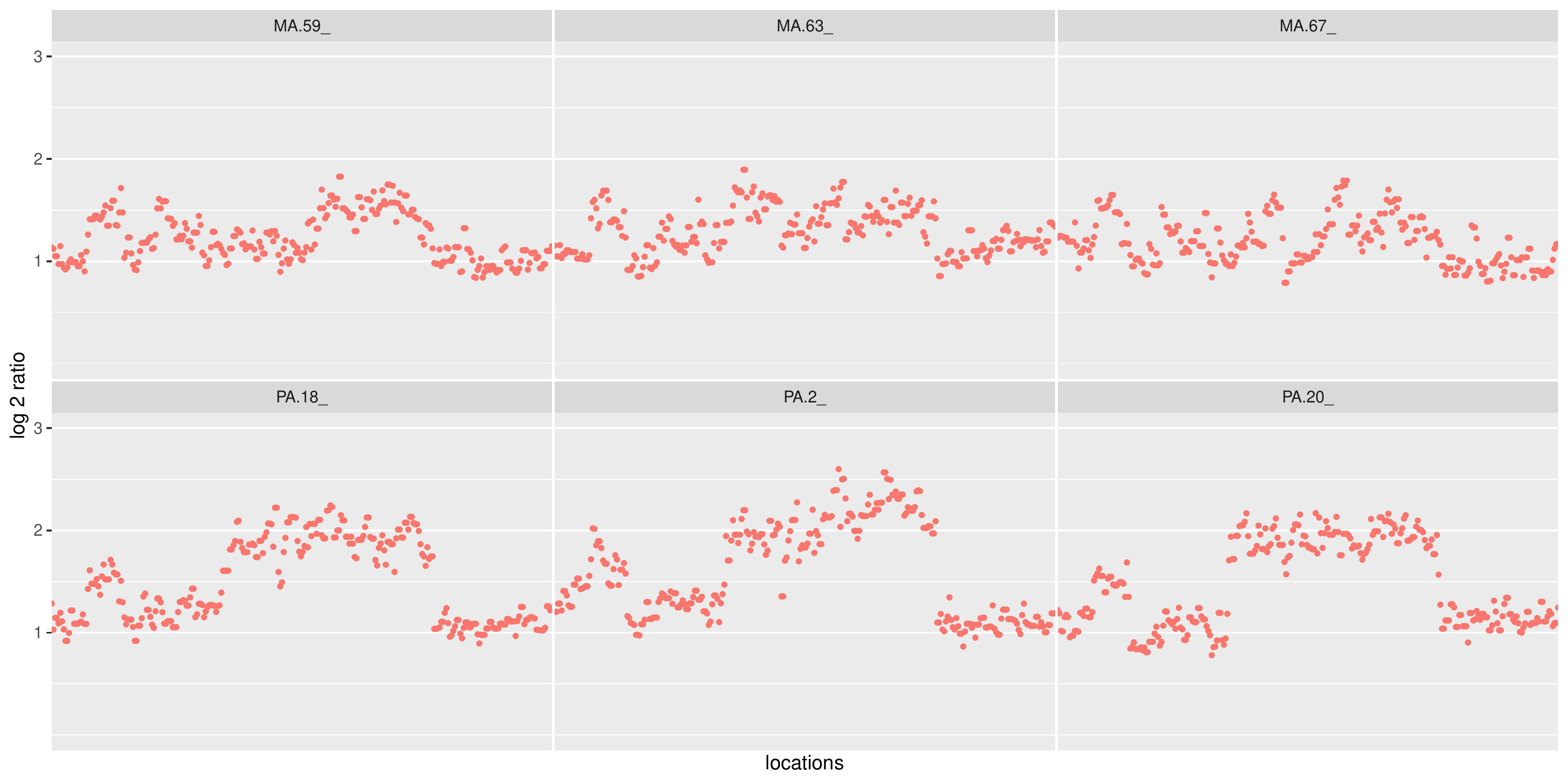}
    \caption{Copy-number data for chromosomes 19, 20 and 21 for six cells, three from primary tumour cells (P) and three from metastatic tumour cells (M). Due to computational cost, read positions were transformed using a median moving window of size five, reducing the total number of genomic bins to 290.}
    \label{fig:app_data}
\end{figure}

Figure \ref{fig:app_data} displays the copy-number data for six cells in our dataset, three from the primary tumour location and three from a liver metastasis location. Our main interest lies in clustering all 45 cells based on their copy-number variations, evaluating whether they form groups according to their tissue of origin and uncovering any novel patterns, if present.

Due to computational cost, we fixed the maximum number of change points ($k^*$) to two, and we applied a median moving window of size five to each data sequence using the zoo package in R \citep{zoo2022} to reduce the number of bins in the data and to handle possible outliers. Considering the transformed data, with 290 locations, we applied our algorithm using two chains of size 10\,000; one was initialized using the clustering result from the K-means method when we set the number of clusters to be two. The other chain was initialized using random cluster assignments; that is, each cell was randomly assigned to one of two clusters. The number of change points for each cluster was set to zero at the beginning of the chains. Additionally, the initial values for the intercepts were selected as the average $\log_2$ ratio copy-number information taken over the cells in each initial cluster, and the sample variances were set as initial values for the variance components. The minimum number of locations in each segment between change points, $w$, was set to 50. Furthermore, convergence was confirmed using the methodology of \cite{Gelman1992} for each chain after the burn-in of half the size of the chains and thinning the remaining samples by selecting every 50th.

\begin{figure}[htbp]
    \centering
    \includegraphics[width = \textwidth]{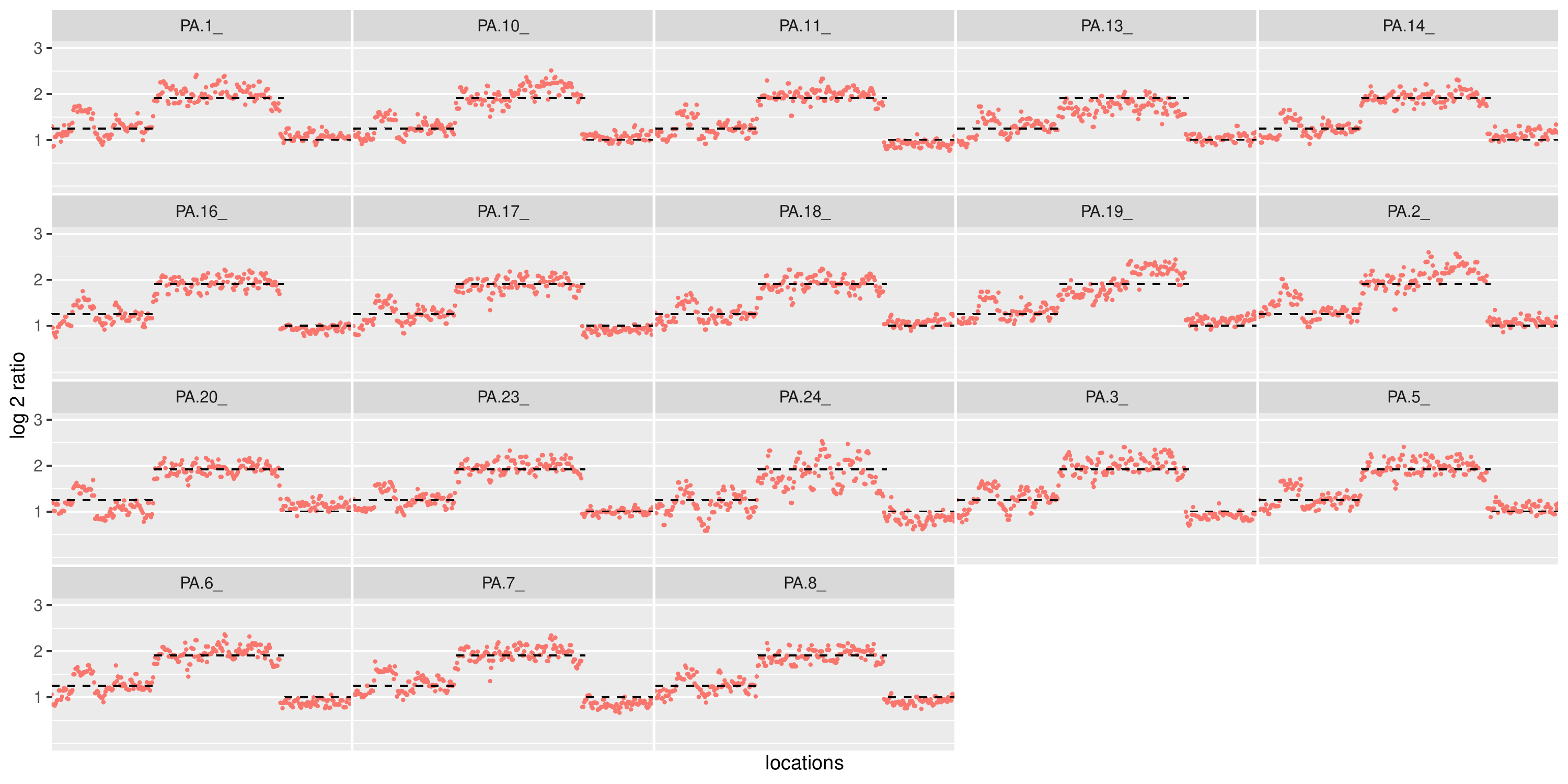}
    \caption{Copy-number data for all 18 cells in Cluster 1 composed of primary tumour cells. Our approach estimated two change points at positions 100 and 226.  The black dashed lines correspond to the mean constant level for each segment between change points.}
    \label{fig:app_cluster1}
\end{figure}

\begin{figure}[htbp]
    \centering
    \includegraphics[width = \textwidth]{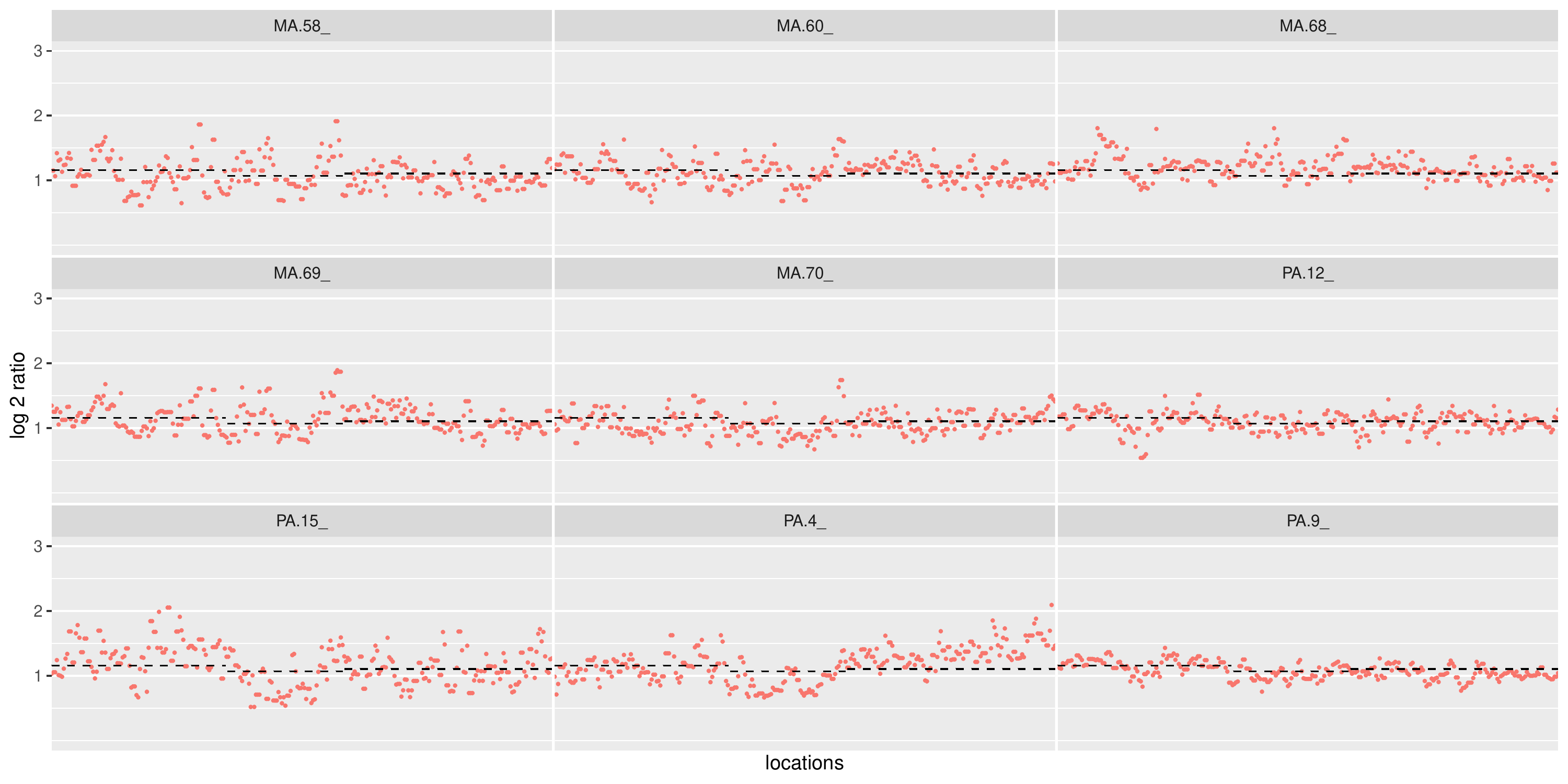}
    \caption{Copy-number data for all nine cells in Cluster 2 composed of tumour cells with copy-number reads around one in all locations. The black dashed lines correspond to the mean constant level for each segment between change points.}
    \label{fig:app_cluster2}
\end{figure}

\begin{figure}[htbp]
    \centering
    \includegraphics[width = \textwidth]{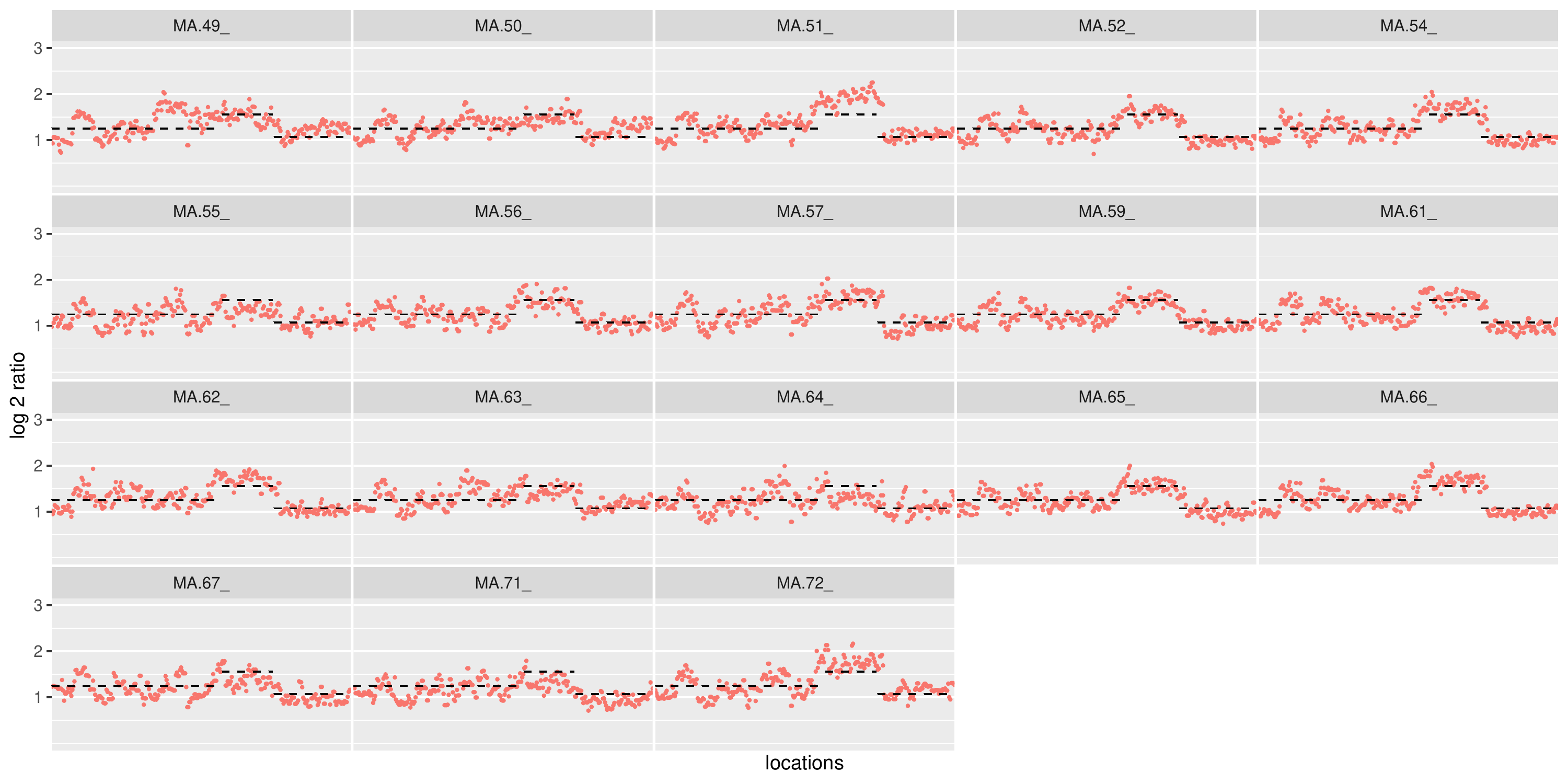}
    \caption{Copy-number data for all 18 cells in Cluster 3 composed of metastatic tumour cells with change points at positions 165 and 215. The black dashed lines correspond to the mean constant level for each segment between change points.}
    \label{fig:app_cluster3}
\end{figure}

Our approach identified three clusters: Cluster 1 is composed of only primary tumour cells with clear change points at bin locations 100 and 226, Cluster 2 with both primary and metastatic tumour cells with log-ratio reads around one for all bins, and Cluster 3 with only metastatic tumour cells with two change points at bin locations 165 and 215, as shown in Figures \ref{fig:app_cluster1}, \ref{fig:app_cluster2} and \ref{fig:app_cluster3}. 

Table \ref{tab:real_data} reports the posterior estimates for the intercepts of each segment between change points for each cluster, where we note that the intercepts for Cluster 2 are not significant since the credible intervals overlap, suggesting as is shown in Figure \ref{fig:app_cluster2} the absence of change points since the log-ratio reads are steady around one for all locations. Interestingly, the cells belonging to Cluster 2 were not considered in the hierarchical clustering analysis in \cite{Leung2017}.  In addition, metastatic and primary tumour single cells were mainly clustered separately, as observed in \cite{Leung2017}. However, \citeauthor{Leung2017} considered all chromosomes when clustering cells and found two clusters for the metastatic tumour cells. Furthermore, they noted that amplifications of chromosomes 3 and 8 distinguished the subpopulations for the metastatic tumour cells. This data was also analyzed in \cite{Safinianaini2020}, where the authors developed a Markov-Chain-based method for clustering copy number data. Analyzing the same data set, \citeauthor{Safinianaini2020} considered the copy-number data for chromosomes 18 to 21 from patient CRC2 to cluster tumour single-cells according to their copy-number profiles. As a result, \citeauthor{Safinianaini2020} identified two clusters of tumour single-cells, separating primary from metastatic single-cells.

\begin{table}[ht]
\centering
\caption{Posterior estimates for the intercepts of each identified cluster. We report the average posterior mean estimates (Average) and standard error (SE) for each intercept, and the $95\%$ credible interval ($95\%$ CI).} 
\label{tab:real_data}
\begin{tabular}{ccrrl}
  \toprule
Cluster & Parameter & Average & SE & $$95\%$$ CI \\ 
  \midrule
 & $\alpha_{1}$ & 1.2536 & 0.0051 & [1.2435, 1.2638] \\ 
  1 & $\alpha_{2}$ & 1.9158 & 0.0044 & [1.9070, 1.9250] \\ 
   & $\alpha_{3}$ & 1.0063 & 0.0061 & [0.9930, 1.0170] \\ \midrule
   & $\alpha_{1}$ & 1.1588 & 0.0127 & [1.1322, 1.1757] \\ 
  2 & $\alpha_{2}$ & 1.0702 & 0.0721 & [1.0161, 1.2301] \\ 
   & $\alpha_{3}$ & 1.1048 & 0.0180 & [1.0655, 1.1235] \\ \midrule
   & $\alpha_{1}$ & 1.2496 & 0.0043 & [1.2417, 1.2568] \\ 
  3 & $\alpha_{2}$ & 1.5595 & 0.0270 & [1.5295, 1.6186] \\ 
   & $\alpha_{3}$ & 1.0732 & 0.0082 & [1.0596, 1.0877] \\ 
   \bottomrule
\end{tabular}
\end{table}

\clearpage
\section{Conclusion}

The results from the simulation scenarios show that our approach can recover the true classification of each data sequence. Furthermore, it is precise in identifying the change points when we vary the number of data sequences and the number of locations. Importantly, the degree of dispersion in the data did not affect our method's performance; we observed satisfactory results in scenarios where the variance components were sampled from inverse-gamma distributions with both small and large averages. Additionally, our method effectively recovers the true underlying data structure in the presence of outliers, demonstrating its robustness. This robustness was evaluated by introducing an outlier in the change-point location for a subset of data sequences from Cluster 1. Using a dataset from Scenario 2 with N = 50 data sequences, we reduced the value of the 19th observation by 10 units in 9\% of the sequences from Cluster 1, causing the first change point for these sequences to shift by one position from its true position. Despite this modification, our method successfully recovered both the true change-point profiles and the correct cluster assignments (results not shown).
Finally, by applying our method to a copy-number single-cell dataset, our approach showed consistent results with \cite{Leung2017}, where we obtained similar clusters for tumour single-cells based on their change-point structures, in which we observed that some cells were clustered according to their tissue of origin. However, the application to the dataset also reveal a novel cluster composed of cells from both primary and metastatic tissue origins, providing new insights into the dataset.

To facilitate the implementation of our method, we have developed the R package BayesCPclust available at \texttt{https://CRAN.R-project.org/package=BayesCPclust}, which to our knowledge, is the first package that addresses the problem of clustering multiple change-point data while simultaneously performing change-point detection.

A limitation of our approach lies in the computational cost, since it requires the calculation of a probability for each possible combination of interval length between change points, which can be computationally expensive when the number of locations increases. To remedy this, in the real data analysis, we calculated the probabilities for a sample of all possible combinations of interval lengths, reducing but not sufficiently the computational cost. In general, as the number of data sequences or locations increases, the average processing time to infer change points and perform clustering analysis for the simulation scenarios also increases, with an average duration between 20 and 30 hours for the scenarios with the highest number of locations (see Table \ref{tab:sc3_time} in the Appendix). Furthermore, we observed similar processing times for the first two scenarios (see Tables \ref{tab:sc1_time} and \ref{tab:sc2_time} in the Appendix), suggesting that the data dispersion had a minimal effect on the computational cost of our algorithm. 

A common issue in Bayesian mixture modeling is that the labels of the clusters can be permuted multiple times over iterations of a Markov Chain Monte Carlo (MCMC) method, such as the Gibbs sampler. This issue, known as label switching, happens since the data likelihood is invariant under the permutation of the labels of the clusters. Solutions for undoing label switching are necessary to perform cluster-specific inference. Thus, various approaches have been proposed to solve this issue \citep{Jasra2005, Papastamoulis2010, Rodriguez2014}. In this paper, we assigned the most frequent set of labels to the sequences of cluster assignments leading to the same clustering. Then, after this correction for label switching, we obtained all the corresponding parameter posterior estimates for each cluster.	

As a possible future work, other Bayesian inference approaches for optimization can be considered, such as variational inference and approximate Bayesian computation methods \citep{Blei2017, Van2021}, which provide functional approximations of posterior distributions and can reduce the computational cost associated with MCMC-based methods. Additionally, the Gibbs sampler developed in this paper can be improved using techniques like blocking, collapsing, and partial collapsing techniques to address slow convergence issues \citep{Park2022}.
\clearpage

\bibliographystyle{apalike}
\bibliography{references.bib}

\clearpage
\section*{Appendix}

\subsection{Gibbs sampler}

\paragraph{Step 1.}

In this step we update the clustering assignments for the observations which depends on
\begin{equation*}
    q_{\ic, 0} = \alpha_0 \sum_{k=0}^{k^*} \sum_{(\ib_1, \dotsc, \ib_{k+1})}H(\ib_1, \dotsc, \ib_{k+1}) \frac{\ib_0}{\ib_1! \dotsm \ib_{k+1}!} \left(\frac{1}{k+1}\right)^{\ib_0} P(\Tcp = k),
\end{equation*}

\noindent where 

\begin{align*}
   H(\ib_1, \dotsc, \ib_{k+1}) &= \int_{\boldsymbol{\alpha}^{(\ic)}} \int_0^\infty f(\boldsymbol{Y}_\ic \mid \boldsymbol{\alpha}^{(\ic)}, \sigma^2_\ic) \pi_3(\sigma^2_\ic) d\sigma^2_\ic d \boldsymbol{\alpha}^{(\ic)} \nonumber \\
   &=  \int_0^\infty \int_{\boldsymbol{\alpha}^{(\ic)}} f(\boldsymbol{Y}_\ic \mid \boldsymbol{\alpha}^{(\ic)}, \sigma^2_\ic) \pi_3(\sigma^2_\ic) d \boldsymbol{\alpha}^{(\ic)} d\sigma^2_\ic,
\end{align*}

\noindent with

\begin{equation*}
    f(\boldsymbol{Y}_\ic \mid \boldsymbol{\alpha}^{(\ic)}, \sigma^2_\ic) = \frac{1}{(2\pi\sigma^2_\ic)^{\Tc/2}} \exp\left\{-\frac{1}{2\sigma^2_\ic}(\boldsymbol{Y}_\ic - X_0\boldsymbol{\alpha}^{(\ic)})^T(\boldsymbol{Y}_\ic - X_0\boldsymbol{\alpha}^{(\ic)})\right\}
\end{equation*}

\noindent and

\begin{equation}
\label{eq:int.alpha}
    \int_{{\boldsymbol{\alpha}}^{(\ic)}}f(\boldsymbol{Y}_\ic \mid \boldsymbol{\alpha}^{(\ic)}, \sigma^2_\ic) d\boldsymbol{\alpha}^{(\ic)}= \int_{{\boldsymbol{\alpha}}^{(\ic)}} \frac{1}{(2\pi\sigma^2_\ic)^{\Tc/2}} \exp\left\{-\frac{1}{2\sigma^2_\ic}(\boldsymbol{Y}_\ic - X_0\boldsymbol{\alpha}^{(\ic)})^T(\boldsymbol{Y}_\ic - X_0\boldsymbol{\alpha}^{(\ic)})\right\}d\boldsymbol{\alpha}^{(\ic)}.
\end{equation}

\noindent We calculate $(\boldsymbol{Y}_\ic - X_0\boldsymbol{\alpha}^{(\ic)})^T(\boldsymbol{Y}_\ic - X_0\boldsymbol{\alpha}^{(\ic)})$ in (\ref{eq:int.alpha}) as:

\begin{align}
\label{eq:square}
(\boldsymbol{Y}_\ic - X_0\boldsymbol{\alpha}^{(\ic)})^T(\boldsymbol{Y}_\ic - X_0\boldsymbol{\alpha}^{(\ic)}) &= \boldsymbol{Y}^T_\ic\boldsymbol{Y}_\ic - \boldsymbol{\alpha}^{T(\ic)}X_0^T\boldsymbol{Y}_\ic  - \boldsymbol{Y}^T_\ic X_0\boldsymbol{\alpha}^{(\ic)} + \boldsymbol{\alpha}^{T(\ic)}X_0^TX_0\boldsymbol{\alpha}^{(\ic)} \nonumber \\
&= \boldsymbol{Y}^T_\ic\boldsymbol{Y}_\ic - 2\boldsymbol{Y}^T_\ic X_0\boldsymbol{\alpha}^{(\ic)} + \boldsymbol{\alpha}^{T(\ic)}X_0^TX_0\boldsymbol{\alpha}^{(\ic)},
\end{align}

\noindent where we can complete the square as follows:

\begin{align}
\label{eq:complete_square}
& (\boldsymbol{\alpha}^{(\ic)} - (X_0^TX_0)^{-1}X_0^T\boldsymbol{Y}_\ic)^T(X_0^TX_0)(\boldsymbol{\alpha}^{(\ic)} - (X_0^TX_0)^{-1}X_0^T\boldsymbol{Y}_\ic) \nonumber \\
&= (\boldsymbol{\alpha}^{T(\ic)} - \boldsymbol{Y}^T_\ic X_0(X_0^TX_0)^{-1})(X_0^TX_0\boldsymbol{\alpha}^{(\ic)} - X_0^TX_0(X_0^TX_0)^{-1}X_0^T\boldsymbol{Y}_\ic ) \nonumber \\
&= \boldsymbol{\alpha}^{T(\ic)}X_0^TX_0\boldsymbol{\alpha}^{(\ic)} - \boldsymbol{\alpha}^{T(\ic)}X_0^T\boldsymbol{Y}_\ic  - \boldsymbol{Y}^T_\ic X_0\boldsymbol{\alpha}^{(\ic)} + \boldsymbol{Y}^T_\ic X_0(X_0^TX_0)^{-1}X_0^T\boldsymbol{Y}_\ic.
\end{align}

\noindent Now, let $V_{\ic} = X_0^TX_0$ and using the result in (\ref{eq:complete_square}) we obtain that Equation (\ref{eq:square}) is equivalent to

\begin{align*}
& \boldsymbol{Y}^T_\ic\boldsymbol{Y}_\ic - 2\boldsymbol{Y}^T_\ic X_0\boldsymbol{\alpha}^{(\ic)} + \boldsymbol{\alpha}^{T(\ic)}X_0^TX_0\boldsymbol{\alpha}^{(\ic)} \\
&=  \boldsymbol{Y}^T_\ic\boldsymbol{Y}_\ic + (\boldsymbol{\alpha}^{(\ic)} - V_{\ic}^{-1}X_0^T\boldsymbol{Y}_\ic)^TV_{\ic}(\boldsymbol{\alpha}^{(\ic)} - V_{\ic}^{-1}X_0^T\boldsymbol{Y}_\ic) - \boldsymbol{Y}^T_\ic X_0V_{\ic}^{-1}X_0^T\boldsymbol{Y}_\ic.
\end{align*}

\noindent Then, we can solve Equation (\ref{eq:int.alpha}) as
\begin{align*}
& \int_{{\boldsymbol{\alpha}}^{(\ic)}}f(\boldsymbol{Y}_\ic \mid \boldsymbol{\alpha}^{(\ic)}, \sigma^2_\ic) d\boldsymbol{\alpha}^{(\ic)}  \nonumber\\
& = \int_{{\boldsymbol{\alpha}}^{(\ic)}} \frac{1}{(2\pi\sigma^2_\ic)^{\Tc/2}} \exp\left\{-\frac{1}{2\sigma^2_\ic}(\boldsymbol{Y}^T_\ic\boldsymbol{Y}_\ic + (\boldsymbol{\alpha}^{(\ic)} - V_{\ic}^{-1}X_0^T\boldsymbol{Y}_\ic)^TV_{\ic}(\boldsymbol{\alpha}^{(\ic)} - V_{\ic}^{-1}X_0^T\boldsymbol{Y}_\ic) - \boldsymbol{Y}^T_\ic X_0V_{\ic}^{-1}X_0^T\boldsymbol{Y}_\ic)\right\}d\boldsymbol{\alpha}^{(\ic)} \nonumber\\
&= \frac{1}{(2\pi\sigma^2_\ic)^{\Tc/2}} \exp\left\{-\frac{1}{2\sigma^2_\ic}(\boldsymbol{Y}^T_\ic\boldsymbol{Y}_\ic - \boldsymbol{Y}^T_\ic X_0V_{\ic}^{-1}X_0^T\boldsymbol{Y}_\ic)\right\} \nonumber \\
&\times \int_{{\boldsymbol{\alpha}}^{(\ic)}} \underbrace{\exp\left\{-\frac{1}{2\sigma^2_\ic}(\boldsymbol{\alpha}^{(\ic)} - V_{\ic}^{-1}X_0^T\boldsymbol{Y}_\ic)^TV_{\ic}(\boldsymbol{\alpha}^{(\ic)} - V_{\ic}^{-1}X_0^T\boldsymbol{Y}_\ic)\right\}}_\text{A}d\boldsymbol{\alpha}^{(\ic)},
\end{align*}

\noindent where $A$ is the kernel of a ($\Tcp_{\ic} + 1$) - variate Normal distribution with mean vector $V_{\ic}^{-1}X_0^T\boldsymbol{Y}_\ic$ and covariance-variance matrix $\sigma^{-2}_{\ic}V_{\ic}^{-1}$. Then, 

\begin{align*}
&\int_{{\boldsymbol{\alpha}}^{(\ic)}}f(\boldsymbol{Y}_\ic \mid \boldsymbol{\alpha}^{(\ic)}, \sigma^2_\ic) d\boldsymbol{\alpha}^{(\ic)} = \frac{1}{(2\pi\sigma^2_\ic)^{\Tc/2}} \exp\left\{-\frac{1}{2\sigma^2_\ic}(\boldsymbol{Y}^T_\ic\boldsymbol{Y}_\ic - \boldsymbol{Y}^T_\ic X_0V_{\ic}^{-1}X_0^T\boldsymbol{Y}_\ic)\right\} \nonumber\\
&\times \frac{(2\pi)^{(\Tcp_\ic + 1)/2}}{(\sigma_\ic^2)^{-(\Tcp_\ic + 1)/2}} \frac{1}{|V_\ic^{-1}|^{-1/2}} \nonumber\\
&= \frac{1}{(2\pi\sigma^2_\ic)^\frac{\Tc-(\Tcp_\ic + 1)}{2}|V_\ic^{-1}|^{-1/2}} \exp\left\{-\frac{1}{2\sigma^2_\ic}(\boldsymbol{Y}^T_\ic\boldsymbol{Y}_\ic - \boldsymbol{Y}^T_\ic X_0V_{\ic}^{-1}X_0^T\boldsymbol{Y}_\ic)\right\}.
\end{align*}

\noindent Let $B = \boldsymbol{Y}^T_\ic\boldsymbol{Y}_\ic - \boldsymbol{Y}^T_\ic X_0V_{\ic}^{-1}X_0^T\boldsymbol{Y}_\ic$, we have that 

\begin{align*}
&\int_{0}^{\infty}\int_{{\boldsymbol{\alpha}}^{(\ic)}} f(\boldsymbol{Y}_\ic \mid \boldsymbol{\alpha}^{(\ic)}, \sigma^2_\ic) \pi_3(\sigma^2_{\ic}) d\boldsymbol{\alpha}^{(\ic)}d\sigma^2_{\ic} \nonumber \\
&=\int_{0}^{\infty} \frac{1}{(2\pi\sigma^2_\ic)^\frac{\Tc-(\Tcp_\ic + 1)}{2}|V_\ic^{-1}|^{-1/2}} \exp\left\{-\frac{1}{2\sigma^2_\ic}B\right\} \times \frac{1}{b^a\Gamma(a)}(\sigma^2_{\ic})^{-a-1} \exp\left\{-\frac{1}{b\sigma^2_\ic}\right\}d\sigma^2_{\ic} \nonumber \\
&=\frac{1}{|V_\ic^{-1}|^{-1/2}(2\pi)^{(\Tc-\Tcp_\ic - 1)/2}}\frac{1}{b^a\Gamma(a)} \int_{0}^{\infty} \underbrace{(\sigma^2_{\ic})^{-\frac{(\Tc-\Tcp_\ic - 1)}{2} - a - 1} \exp\left\{-\frac{1}{\sigma^2_\ic}\left(\frac{B}{2} + \frac{1}{b}\right)\right\}}_{E}d\sigma^2_{\ic},
\end{align*}

\noindent where $E$ is the kernel of an inverse-gamma distribution with parameters $a_1 = \frac{(\Tc-\Tcp_\ic - 1)}{2} + a$ and $b_1 =  \frac{B}{2} + \frac{1}{b}$. Then, 

\begin{align*}
&\int_{0}^{\infty}\int_{{\boldsymbol{\alpha}}^{(\ic)}} f(\boldsymbol{Y}_\ic \mid \boldsymbol{\alpha}^{(\ic)}, \sigma^2_\ic) \pi_3(\sigma^2_{\ic})d\boldsymbol{\alpha}^{(\ic)} d\sigma^2_{\ic} \nonumber \\
&=\frac{1}{|V_\ic^{-1}|^{-1/2}(2\pi)^{(\Tc-\Tcp_\ic - 1)/2}}\frac{1}{b^a\Gamma(a)} \frac{\Gamma(\frac{\Tc-\Tcp_\ic - 1}{2} + a)}{\left(\frac{B}{2} + \frac{1}{b}\right)^{\frac{(\Tc-\Tcp_\ic - 1)}{2} + a}}.
\end{align*}

\noindent Therefore we have that 

\begin{equation*}
H(\ib_1, \dotsc, \ib_{k+1}) = \frac{1}{|V_\ic|^{1/2}(2\pi)^{(\Tc-\Tcp_\ic - 1)/2}}\frac{1}{b^a\Gamma(a)} \frac{\Gamma(\frac{\Tc-\Tcp_\ic - 1}{2} + a)}{\left(\frac{B}{2} + \frac{1}{b}\right)^{\frac{(\Tc-\Tcp_\ic - 1)}{2} + a}}, 
\end{equation*}

\noindent implying that
\begin{align}
\label{eq:qn0}
    q_{\ic, 0} = \alpha_0 \sum_{k=0}^{k^*} \sum_{(\ib_1, \dotsc, \ib_{k+1})} \frac{1}{|V_\ic|^{1/2}(2\pi)^{(\Tc-\Tcp_\ic - 1)/2}}\frac{1}{b^a\Gamma(a)} \frac{\Gamma(\frac{\Tc-\Tcp_\ic - 1}{2} + a)}{\left(\frac{B}{2} + \frac{1}{b}\right)^{\frac{(\Tc-\Tcp_\ic - 1)}{2} + a}}&\times \\ \nonumber
    \frac{\ib_0}{\ib_1! \dotsm \ib_{k+1}!} \left(\frac{1}{k+1}\right)^{\ib_0} P(\Tcp = k).
\end{align}

\noindent Now, for $q_{\ic,j}$ we first define $\ell(\boldsymbol{Y}_\ic \mid \bt_{(\icluster)})$ as:

\begin{align*}
&\ell(\boldsymbol{Y}_\ic \mid \bt_{(\icluster)}) = \int_0^\infty f(\boldsymbol{Y}_\ic \mid \bt_{(\icluster)}, \sigma_\ic^2) \pi_3(\sigma^2_\ic) d\sigma_\ic^2  \nonumber \\
&=\int_0^\infty \frac{1}{(2\pi\sigma^2_\ic)^{\Tc/2}} \exp\left\{-\frac{1}{2\sigma^2_\ic}(\boldsymbol{Y}_\ic - X_0\boldsymbol{\alpha}^{(\ic)})^T(\boldsymbol{Y}_\ic - X_0\boldsymbol{\alpha}^{(\ic)})\right\} \nonumber \\
& \times \frac{1}{b^a\Gamma(a)}(\sigma^2_{\ic})^{-a-1} \exp\left\{-\frac{1}{b\sigma^2_\ic}\right\}d\sigma^2_{\ic} \nonumber\\
&=\int_0^\infty \frac{1}{b^a(2\pi)^{\Tc/2}\Gamma(a)}(\sigma^2_{\ic})^{-\frac{\Tc}{2}-a-1} \exp\left\{-\frac{1}{\sigma^2_\ic}\left(\frac{\boldsymbol{Y}_\ic - X_0\boldsymbol{\alpha}^{(\ic)})^T(\boldsymbol{Y}_\ic - X_0\boldsymbol{\alpha}^{(\ic)}}{2} + \frac{1}{b}\right) \right\}d\sigma^2_{\ic} \nonumber\\
&= \frac{\Gamma\left(\frac{\Tc}{2}+a\right)}{b^a(2\pi)^{\Tc/2}\Gamma(a)}\left(\frac{\boldsymbol{Y}_\ic - X_0\boldsymbol{\alpha}^{(\ic)})^T(\boldsymbol{Y}_\ic - X_0\boldsymbol{\alpha}^{(\ic)}}{2} + \frac{1}{b}\right)^{-(\Tc/2 + a)}.
\end{align*}

\noindent Then, 

\begin{equation}
\label{eq:qnj}
q_{\ic,j} = \frac{\Gamma\left(\frac{\Tc}{2}+a\right)}{b^a(2\pi)^{\Tc/2}\Gamma(a)}\left(\frac{(\boldsymbol{Y}_\ic - X_0\boldsymbol{\alpha}^{(\ic)})^T(\boldsymbol{Y}_\ic - X_0\boldsymbol{\alpha}^{(\ic)})}{2} + \frac{1}{b}\right)^{-(\Tc/2 + a)}.  
\end{equation}

In practice, in Step 1, we start with a Bernoulli experiment, generating 0 and 1 with probabilities $p = q_{\ic,0} / q_{\ic,0} + \sum_{j =1}^{\Tcluster} \ic_j q_{\ic,j}$ and $1 - p$. If 0 results, a new $\bt_s$ is generated from $G^*(d\bt_\ic)$, $\Tcluster$ is increased to $\Tcluster$ + 1. If 1 results, the existing cluster label $j$ is sampled with probability $p_j = \ic_j q_{\ic,j}/ \sum_{j=1}^{\Tcluster} \ic_j q_{\ic,j}$ for $j = 1, 2, 3, \dotsc, \Tcluster$.  If $j^*$ is sampled, $\bt_\ic$ is set to $\bt_{j^*}$.

\paragraph{Step 2.}

\begin{equation*}
P(\sigma^2_\ic \mid \bt_\ic, \boldsymbol{Y}_\ic) \propto f(\boldsymbol{Y}_\ic \mid \bt_\ic, \sigma_\ic^2) \pi_3(\sigma^2_\ic)    
\end{equation*}

\begin{align*}
&P(\sigma^2_\ic \mid \dotso) \propto   \frac{1}{(2\pi\sigma^2_\ic)^{\Tc/2}} \exp\left\{-\frac{1}{2\sigma^2_\ic}(\boldsymbol{Y}_\ic - X_0\boldsymbol{\alpha}^{(\ic)})^T(\boldsymbol{Y}_\ic - X_0\boldsymbol{\alpha}^{(\ic)})\right\} \nonumber\\ 
&\times \frac{}{b^a\Gamma(a)}(\sigma^2_{\ic})^{-a-1} \exp\left\{-\frac{1}{b\sigma^2_\ic}\right\} \nonumber\\
&\propto (\sigma^2_{\ic})^{-\frac{\Tc}{2}-a-1} \exp\left\{-\frac{1}{\sigma^2_\ic}\left(\frac{\boldsymbol{Y}_\ic - X_0\boldsymbol{\alpha}^{(\ic)})^T(\boldsymbol{Y}_\ic - X_0\boldsymbol{\alpha}^{(\ic)}}{2} + \frac{1}{b}\right) \right\} \nonumber\\ 
&\sim \mathrm{inverse-gamma}\left(\frac{\Tc}{2}+a,\frac{(\boldsymbol{Y}_\ic - X_0\boldsymbol{\alpha}^{(\ic)})^T(\boldsymbol{Y}_\ic - X_0\boldsymbol{\alpha}^{(\ic)})}{2} + \frac{1}{b}\right).
\end{align*}

\paragraph{Step 3.}

In this step, we calculate the posterior marginal probability function of $\Tcp_\icluster$ which depends on
\begin{equation}
\label{eq:v_def1}
v(\ib_1, \dotsc, \ib_{k+1}) = \exp{(\tilde{H}(\ib_1, \dotsc, \ib_{k+1}))}\frac{\Gamma(\ib_0 + 1)}{\prod_{l=1}^{k+1}\Gamma(\ib_l + 1)}\left(\frac{1}{k+1}\right)^{\ib_0},
\end{equation}

\noindent where

\begin{equation*}
\tilde{H}(\ib_1, \dotsc, \ib_{k+1}) = \log\int_{\mathbb{R}^{k+1}} \prod_{\ic \in C_\icluster} f(\boldsymbol{Y}_\ic \mid \boldsymbol{\alpha}^{(\icluster)}, \sigma^2_\ic) d\boldsymbol{\alpha}^{(\icluster)}.    
\end{equation*}

\noindent We can write $\prod_{\ic \in C_\icluster} f(\boldsymbol{Y}_\ic \mid \boldsymbol{\alpha}^{(\icluster)}, \sigma^2_\ic)$ as:

\begin{align*}
&\prod_{\ic \in C_\icluster} f(\boldsymbol{Y}_\ic \mid \boldsymbol{\alpha}^{(\icluster)}, \sigma_{\icluster}^2) = \prod_{\ic \in C_\icluster} \frac{1}{(2\pi\sigma^2_\ic)^{\Tb/2}} \exp\left\{-\frac{1}{2\sigma^2_\ic}(\boldsymbol{Y}_\ic - X_{0,\icluster}\boldsymbol{\alpha}^{(\icluster)})^T(\boldsymbol{Y}_\ic - X_{0,\icluster}\boldsymbol{\alpha}^{(\icluster)})\right\} \nonumber\\
&= \frac{1}{(2\pi)^{|C_\icluster|\Tb/2}}\prod_{\ic \in C_\icluster}({\sigma^2_\ic)^{-\Tb/2}} \exp\left\{-\sum_{\ic \in C_\icluster}\frac{\sigma^{-2}_\ic}{2}(\boldsymbol{Y}_\ic - X_{0,\icluster}\boldsymbol{\alpha}^{(\icluster)})^T(\boldsymbol{Y}_\ic - X_{0,\icluster}\boldsymbol{\alpha}^{(\icluster)})\right\} \nonumber\\
&=\frac{1}{(2\pi)^{|C_\icluster|\Tb/2}}\prod_{\ic \in C_\icluster} ({\sigma^2_\ic)^{-\Tb/2}} \exp\left\{-\sum_{\ic \in C_\icluster}\frac{\sigma^{-2}_\ic}{2}(\boldsymbol{Y}^T_\ic\boldsymbol{Y}_\ic  - \boldsymbol{Y}^T_\ic X_{0,\icluster}V^{-1}_\icluster X_{0,\icluster}^T\boldsymbol{Y}_\ic)\right\} \times  \nonumber\\
& \exp\left\{-\sum_{\ic \in C_\icluster}\frac{\sigma^{-2}_\ic}{2}(\boldsymbol{\alpha}^{(\icluster)} - V_{\icluster}^{-1}X_{0,\icluster}^T\boldsymbol{Y}_\ic)^TV_{\icluster}(\boldsymbol{\alpha}^{(\icluster)} - V_{\icluster}^{-1}X_{0,\icluster}^T\boldsymbol{Y}_\ic) \right\}.
\end{align*}

\noindent Now, using the same approach as in Equation (\ref{eq:square}) with $V_\icluster = X_{0,\icluster}^TX_{0,\icluster}$, we obtain:

\begin{align}
\label{eq:v_def2}
&\tilde{H}(\ib_1, \dotsc, \ib_{k+1}) = \log\int_{\mathbb{R}^{k+1}} \prod_{\ic \in C_\icluster} f(\boldsymbol{Y}_\ic \mid \boldsymbol{\alpha}^{(\icluster)}, \sigma_{\ic}^2) d\boldsymbol{\alpha}^{(\icluster)} \nonumber\\
&= -\frac{|C_\icluster|\Tb - k - 1}{2} \log(2\pi) - \frac{\Tb}{2} \left(\sum_{\ic \in C_\icluster}{\log(\sigma^{2}_\ic)}\right) - \frac{1}{2}\sum_{\ic \in C_\icluster}\sigma^{-2}_\ic(\boldsymbol{Y}^T_\ic\boldsymbol{Y}_\ic  - \boldsymbol{Y}^T_\ic X_{0,\icluster}V^{-1}_\icluster X_{0,\icluster}^T\boldsymbol{Y}_\ic) \nonumber \\
& -\frac{k+1}{2} \log\left(\sum_{\ic \in C_\icluster}{\sigma^{-2}_\ic}\right) -\frac{1}{2} \log|V_r|.
\end{align}

Then, after updating the interval lengths, $\boldsymbol{\tau}_\icluster$, for cluster $\icluster$ we obtain the updates for the intercepts $\boldsymbol{\alpha}_\icluster$ for cluster $\icluster$ using the following full conditional distribution.

\begin{align*}
P(\boldsymbol{\alpha}^{(\icluster)} \mid \boldsymbol{\tau}_\icluster, \Tcp_\icluster, \boldsymbol{Y}_\icluster) &\propto f(\boldsymbol{Y} \mid \boldsymbol{\alpha}^{(\icluster)}, \boldsymbol{\sigma}^2)  \pi(\boldsymbol{\alpha}^{(\icluster)}) \nonumber\\
&=\prod_{\ic \in C_\icluster} f(\boldsymbol{Y}_\ic \mid \boldsymbol{\alpha}^{(\icluster)}, \sigma_{\ic}^2) \times 1 \nonumber\\
& \propto \exp\left\{-\sum_{\ic \in C_\icluster}\frac{\sigma^{-2}_\ic}{2}(\boldsymbol{\alpha}^{(\icluster)} - V_{\icluster}^{-1}X_{0,\icluster}^T\boldsymbol{Y}_\ic)^TV_{\icluster}(\boldsymbol{\alpha}^{(\icluster)} - V_{\icluster}^{-1}X_{0,\icluster}^T\boldsymbol{Y}_\ic) \right\} \nonumber\\
&\sim \mathrm{Normal}\left(V^{-1}_\icluster X_{0,\icluster}^T\boldsymbol{Y}_\ic, V^{-1}_\icluster \sum_{\ic \in C_\icluster}\sigma^{-2}_\ic\right).
\end{align*}

\subsection{Additional simulation results}

\begin{table}[ht]
\centering
\caption{Simulation scenario 1. Average processing time taken over 96 datasets randomly generated from our proposed model with 5000 iterations when we vary the number of data sequences with small data dispersion.} 
\label{tab:sc1_time}
\begin{tabular}{lr}
  \toprule
$\Tc$ & Average processing time (minutes) \\ 
  \midrule
10 & 157.2476 \\ 
  25 & 291.2428 \\ 
  50 & 513.2888 \\  
   \bottomrule
\end{tabular}
\end{table}

\begin{table}[ht]
\centering
\caption{Simulation scenario 2. Average processing time taken over 96 datasets randomly generated from our proposed model with 5000 iterations when we vary the number of data sequences with higher data dispersion.} 
\label{tab:sc2_time}
\begin{tabular}{lr}
  \toprule
$\Tc$ & Average processing time (minutes) \\ 
  \midrule
10 & 153.6956 \\ 
  25 & 282.8196 \\ 
  50 & 500.9338 \\ 
   \bottomrule
\end{tabular}
\end{table}

\begin{table}[ht]
\centering
\caption{Simulation scenario 3. Average processing time taken over 96 datasets randomly generated from our proposed model with 5000 iterations when we vary the number of locations.} 
\label{tab:sc3_time}
\begin{tabular}{lr}
  \toprule
$\Tb$ & Average processing time (minutes) \\ 
  \midrule
50 & 283.5759 \\ 
100 & 1715.0926 \\ 
200 & 1274.1529 \\ 
   \bottomrule
\end{tabular}
\end{table}

\end{document}